\documentclass[twocolumn,times]{revtex4}
\usepackage{epsfig,graphics,amssymb,amsmath,rotating,color}

\def\n{\mathbf{n}}
\def\t{\mathbf{t}}

\def\X{\mathbf{X}}

\def\R{\mathbf{R}}

\def\b{\mathbf{b}}

\def\I{\mathbf{I}}

\def\H{\mathbf{H}}
\def\v{\mathbf{v}}
\def\f{\mathbf{f}}

\def\bs{s'}

\def\U{\mathbf{U}}

\def\varepsilon{\epsilon}

\begin{document}

\title{Hydrodynamics of the double-wave structure of insect spermatozoa flagella}
\author{On Shun Pak}
\author{Saverio E. Spagnolie}
\author{Eric Lauga}
\affiliation{Department of Mechanical and Aerospace Engineering, University of California San Diego, 9500 Gilman Drive, La Jolla CA 92093-0411.}
%\date{\today}

\begin{abstract}
In addition to conventional planar and helical flagellar waves, insect sperm flagella have also been observed to display a double-wave structure characterized by the presence of two superimposed helical waves. In this paper, we present a hydrodynamic investigation of the locomotion of insect spermatozoa exhibiting the double-wave structure, idealized here as superhelical waves. Resolving the hydrodynamic interactions with a non-local slender body theory, we predict the swimming kinematics of these superhelical swimmers based on experimentally collected geometric and kinematic data. Our consideration provides insight into the relative contributions of the major and minor helical waves to swimming; namely, propulsion is due primarily to the minor wave, with negligible contribution from the major wave. We also explore the dependence of the propulsion speed on geometric and kinematic parameters, revealing counter-intuitive results, particularly for the case when the minor and major helical structures are of opposite chirality.
 \end{abstract}
\maketitle

\section{\label{sec:introduction}Introduction}

Locomotion in fluids is ubiquitous in nature, with examples spanning a wide range in size from bacterial motility to the swimming of whales. It plays fundamental roles throughout the lives of animals in such endeavors as predation and finding a mate for reproduction \cite{biewener}.  Lying along the interface between biology and fluid dynamics, biological locomotion at small scales has received substantial attention from biologists and engineers in recent years \cite{fauci3,lauga2}.

The physics governing locomotion in fluids is very different for microscopic organisms (e.g.~bacteria, spermatozoa) and macroscopic organisms (e.g.~fish, humans). The dramatic difference is due to the competition between inertial and viscous effects in the fluid medium. The Reynolds number, $Re=U\,L/\nu$, (with $U$ and $L$ characteristic velocity and length scales, and $\nu$ the kinematic viscosity) is a dimensionless parameter which measures the relative importance of the inertial forces to viscous forces in a fluid. Locomotion of larger animals in fluids takes place at moderate to large Reynolds numbers, where inertial forces dominate. At this scale, swimming and flying are generally accomplished by imparting momentum into the fluid opposite the direction of locomotion. Microorganisms meanwhile inhabit in a world of low Reynolds numbers, where inertia plays a negligible role and viscous damping is paramount. The Reynolds number ranges from $10^{-6}$ for bacteria to $10^{-2}$ for spermatozoa \cite{brennen}. The absence of inertia imposes stringent constraints on a microorganism's locomotive capabilities. 

Many microorganisms propel themselves by propagating travelling waves along one or many slender flagella \cite{brennen}. The motility features of these flagella depend on the cell type, either prokaryotic (cells without a nucleus) or eukaryotic (cells with nuclei). The flagella of prokaryotic bacteria, such as those utilized by \textit{Esherichia coli}, are helical in shape and are passively rotated at their base by a motor embedded in the cell wall. The rotation propagates an apparent helical wave from the sperm head to the distal end of the flagellum, propelling the cell in the opposite direction. Eukaryotic flagella exhibit a different internal structure,  called an axoneme,  which is composed of microtubules, proteins and protein complexes such as dynein molecular motors. The dynein arms convert chemical energy contained in ATP into mechanical energy, inducing active relative sliding between the microtubules, which in turn leads to bending deformations which propagate along the flagellum. A common structure of the axoneme has a ring of nine microtubule doublets spaced around the circumference and two additional central microtubules (the so-called 9+2 axoneme). Other variations of the axonemal structure have also been observed \cite{werner08}.

\begin{figure}[!]
\begin{center}
\includegraphics[width=0.45\textwidth]{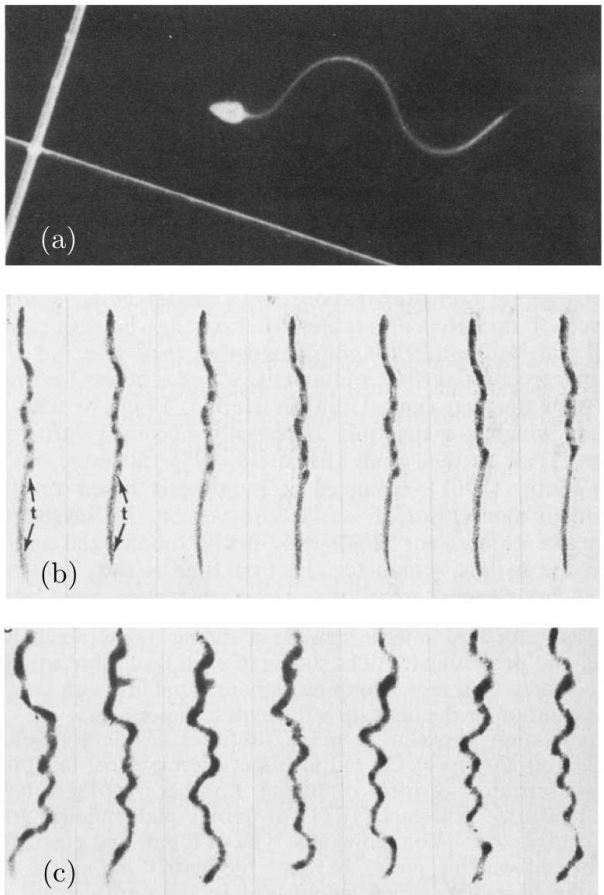}
\end{center}
\caption{\footnotesize  A hierarchy of the complexity of flagellar beating pattern observed in eukaryotic cells. (a) Planar-wave pattern in sea-urchin spermatozoa flagella \cite{rikmenspoel85}; (b) helical-wave pattern in \textit{Gryllotalpa gryllotalpa} \cite{baccetti72}; (c) double-wave pattern in \textit{Haematopinus suis} \cite{baccetti72}.  All images reproduced with permission; (a) from R. Rikmenspoel and C. A. Isles, \textit{Biophys. J.}, 47, 395--410, 1985, copyright 1985 Elsevier; (b) \& (c) from B. Baccetti, \textit{Adv. Insect Physiol.}, 9, 315--397, 1972, copyright 1972 Elsevier.}
\label{fig:hierarchy}
\end{figure}

\begin{figure*}
\begin{center}
\includegraphics[width=0.6\textwidth]{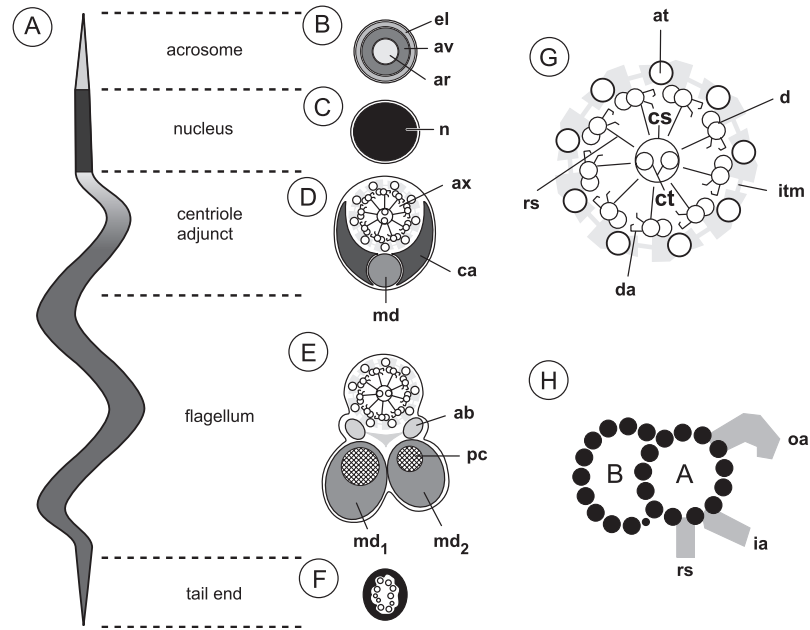}
\end{center}
\caption{\footnotesize Schematic representation of the ground plan of an pterygote insect flagellosperm and its ultrastructure. (A) A typical filiform insect spermatozoon. Although not easily visible from the outside it can be divided into five distinct parts: acrosome, nucleus, centriole adjunct, flagellum, and tail end. (B) Cross section of the acrosome showing its trilayered arrangement of an inner acrosomal rod (ar), an acrosomal vesicle (av) and an outer extra acrosomal layer (el). (C) Cross section of the nucleus (n) showing condensed chromatin. (D) Cross section of the posterior centriole adjunct region. This part of the spermatozoon is characterized by the electron dense centriole adjunct material (ca), often surrounding the anterior part of the axoneme (ax) and the tip of one of the mitochondrial derivatives (md). (E) Cross section through a representative segment of the flagellum. In addition to the axoneme, two accessory bodies (ab) and the two mitochondrial derivatives of often different size (md1, md2) can be seen. The mitochondrial derivatives typically bear paracrytalline inclusions (pc). (F) Cross section through the tail end showing dissociated axonemal tubules. (G) Cross sectional representation of a typical 9]9]2 insect sperm axoneme. nine microtubule doublets (d) with associated dynein arms (da) and radial spokes (rs) are connected to two central microtubules (ct) via the central sheath (cs). The doublets are in turn surrounded by nine accessory tubules (at). Accessory tubules and doublets are linked together by intertubular material (itm). (H) Schematic cross sectional drawing of an axonemal doublet showing the protofilament arrangement of the A and B subtubules. The radial spoke (rs), the inner dynein arm (ia) and the outer dynein arm (oa) are attached to the A subtubule. This figure and caption are reproduced with permission from M. Werner and L. W. Simmons, \textit{Biol. Rev.}, 83, 191--208, 2008, copyright 2008 John Wiley and Sons.}
\label{fig:axoneme}
\end{figure*}

Generally, three levels of complexity in undulatory beat patterns are observed in eukaryotic flagella \cite{baccetti72, werner08}, following a hierarchy in the structure of the axoneme: (1) the lowest in the hierarchy is a simple planar beating pattern, as in human and sea-urchin spermatozoa flagella (Fig.~\ref{fig:hierarchy}a), with the common 9+2 axoneme structure; (2) a more complicated three-dimensional helical beating pattern is exhibited by some insect spermatozoa with a 9+9+2 axoneme, as in \textit{Gryllotalpa gryllotalpa} (Fig.~\ref{fig:hierarchy}b); (3) the highest level of complexity is a double-wave pattern observed in some insect spermatozoa with a 9+9+2 axoneme and accessory bodies endowed with ATPase activity, as in \textit{Haematopinus suis} (Fig.~\ref{fig:hierarchy}c). A vast diversity in sperm structure is found in insects \cite{jamieson99}, and the hierarchy described is also observed even just within the realm of insect spermatozoa flagella  \cite{baccetti72, werner08}. Fig.~\ref{fig:axoneme} shows a schematic diagram of an pterygote insect flagellosperm and its ultrastructure (reproduced with permission from Ref.~\cite{werner08}). Similar to the 9+2 axoneme observed in cilia and flagella of many plant and animal cells, the central core of the insect sperm axoneme is composed of two central microtubules surrounded by a ring of nine microtubule doublets. However, the ring of nine microtubule doublets is surrounded by another nine accessory tubules, forming the characteristic 9+9+2 arrangement of the insect sperm axoneme. In addition to the more complicated microtubule arrangement, two prominent features of inset spermatozoa flagella are the mitochondrial derivatives and accessory bodies running along the axoneme (see Ref.~\cite{werner08} for a thorough review of insect sperm structure).
%%%%%%%%%%%%%%%%%%%%%%%%%%%%%%%%%%%

Although the structure of many different spermatozoa has been examined, the rapid and divergent evolution in sperm morphology is not well understood \cite{werner08, morrow}. Hydrodynamic considerations of the relationship between flagellar morphology and functional parameters such as the swimming speed may provide useful information for explaining the evolutionary divergence. Due to its intricate nature, the double-wave structure is less well explored than wave types sitting lower in the hierarchy. Relevant studies on the planar and helical wave structures are abundant and well developed (see the classical and recent reviews \cite{lauga2,brennen,fauci3}), but we are not aware of any hydrodynamic studies dedicated to the double-wave structure. Here we present a hydrodynamic study on the motility of insect spermatozoa exhibiting a double-wave beat pattern.

\begin{figure*}[!]
\begin{center}
\includegraphics[width=0.7\textwidth]{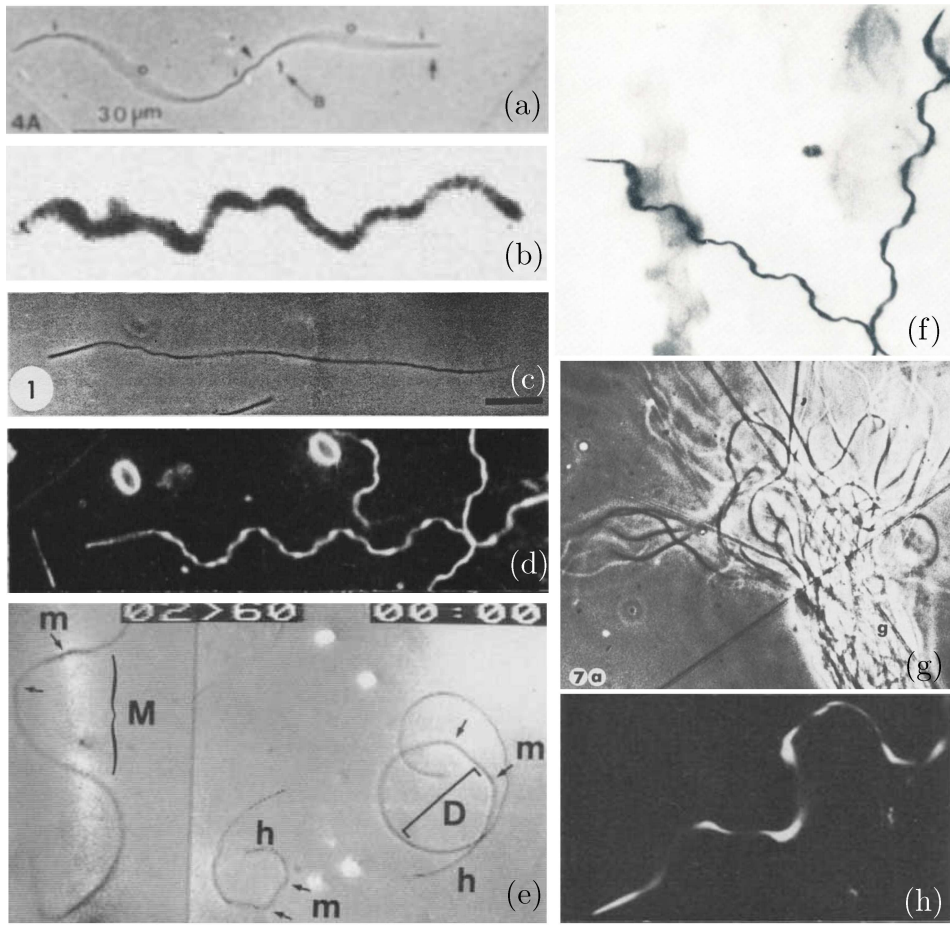}
\end{center}
\caption{\footnotesize  The double-wave beating pattern observed in insect spermatozoa flagella of different species: (a) \textit{Megaselia scalaris} \cite{curtis91}; (b) \textit{Haematopinus suis} \cite{baccetti72}; (c) \textit{Culicoides melleus} \cite{linley79}; (d) \textit{Tenebrio molitor} \cite{baccetti73a}; (e) \textit{Drosophila obscura} \cite{bressac91}; (f) \textit{Tenebrio molitor} \cite{philips74}; (g) \textit{Aedes notoscriptus} \cite{swan81}; (h) \textit{Bacillus rossius} \cite{baccetti73b}. All images reproduced with permission;  (a) from S. K. Curtis and D. B. Benner, \textit{J. Morphol.}, 210, 85--99, 1991, copyright 1991 John Wiley and Sons; (b) from B. Baccetti, \textit{Adv. Insect Physiol.}, 9, 315--397, 1972, copyright 1972 Elsevier; (c) from J. R. Linley, \textit{Entomol. Exp. Appl.}, 26, 85--96, 1979, copyright 1979 John Wiley and Sons; (d) from B. Baccetti, A. G. Burrini, R. Dallai, F. Giusti, M. Mazzini, T. Renieri, F. Rosati, and G. Selmi, \textit{J. Mechanochem. Cell Motil.}, 2, 149--161, 1973, copyright 1973 Plenum Publishing Corporation, with kind permission from Springer Science+Business Media B.V; (e) from C. Bressac, D. Joly, J. Devaux, C. Serres, D. Feneus, and D. Lachaise, \textit{Cell Motil. Cytoskel.}, 19, 269--274, 1991, copyright 1991 John Wiley and Sons; (f) from D. M. Philips, in M. A. Sleigh, editor, \textit{Cilia and Flagella}, 379--402, 1974, copyright 1974 John Wiley and Sons; (g) from M. A. Swan, \textit{Gamete Res.}, 4, 241--250, 1981, copyright 1981 John Wiley and Sons; (h) from B. Baccetti, A. G. Burrini, R. Dallai, V. Pallini, P. Periti, F. Piantelli, F. Rosati, and G. Selmi,  \textit{J. Ultrastruct. Res.}, 44, 1--73, 1973, copyright 1973 Elsevier.}
\label{fig:doublewave}
\end{figure*}

The double-wave pattern is characterized by the simultaneous presence of two kinds of waves, a minor wave with small amplitude and high frequency superimposed on a major helical wave of large amplitude and low frequency. The minor wave has also been observed to be approximately helical \cite{werner02}, and the combined activity of the two is described as a double-helical beating pattern \cite{werner08}. The double-wave structure was first observed in \textit{Tenebrio molitor} and \textit{Bacillus rossius} by Baccetti \textit{et al.} \cite{baccetti73a,baccetti73b}, and was also later found in \textit{Lygaeus} \cite{philips74}, \textit{Culicoides melleus} \cite{linley79}, \textit{Aedes notoscriptus} \cite{swan81}, \textit{Ceratitis capitata} \cite{baccetti89}, \textit{Drosophila obscura} \cite{bressac91}, \textit{Megaselia scalaris} \cite{curtis91}, and more recently in \textit{Aleochara curtula} \cite{werner02} and \textit{Drusilla canaliculata} \cite{werner07}. Figure~\ref{fig:doublewave} compiles a collection of images of spermatozoa exhibiting the double-wave structure. Werner \& Simmons \cite{werner08} have presented a thorough review of this complex structure in insect spermatozoa. 

Most studies on insect spermatozoa focus on the sperm ultrastructure, and there are very few studies on insect sperm motility \cite{werner08}. Many important geometric and kinematic data required for hydrodynamic modeling of the double-wave structure are unavailable. In particular, we are not aware of any information about the chirality of the minor helical structure relative to the major helical structure. The generation and propagation mechanism of the double-wave is also not yet fully understood. Baccetti \textit{et al.} \cite{baccetti73a, baccetti73b} have suggested that the accessory bodies and the axoneme are responsible for the major and minor waves, respectively, whereas Swan \cite{swan81} has stated that the major wave may be due to the sliding of the accessory tubule against the axonemal doublets. More recently, Werner \textit{et al.} \cite{werner02} have proposed a completely different line of thought, suggesting that the major wave is not in fact a real wave but a static helical structure formed due to the coupling of static forces of the axoneme, mitochondrial derivatives, and plasma membrane. The apparent propagation of the major wave could be due to the passive rolling of the entire cell and might in fact be mistaken for an active, propagating wave under the microscope. It has therefore been suggested that the sperm motility is caused solely by the minor wave. The relative extent of the contribution of the major and minor waves to propulsion is thus still an open question \cite{linley79}. With the hydrodynamic study presented in this paper, we hope to provide physical insights on these unresolved problems.

The structure of this paper is as follows. We idealize the double-wave structure as the propagation of superhelical waves and model the hydrodynamics using non-local slender body theory in \S\ref{sec:formulation}. In \S\ref{sec:results}, we present the computed hydrodynamic performance of spermatozoa of different species and compare the predictions with available experimental data (\S\ref{sec:performance}). The features of superhelical swimming are illustrated by a specific model organism, namely the spermatozoa of \textit{Culicoides melleus} (\S\ref{sec:modelproblem}). We then investigate the effects of kinematic and geometric parameters on the propulsion performance of a superhelical swimmer (\S\ref{sec:parametric}). Finally, the limitations of the present study and directions for future work are discussed in \S\ref{sec:discussion}.

\section{\label{sec:formulation}Materials and Methods}
\subsection{Idealized double-wave structure: superhelical swimmers}

The experimentally observed double-wave structure of insect spermatozoa is mathematically idealized in this paper as a superhelix (a small helix itself coiled into a larger helix); we refer to the helical structure with the larger wavelength as the major helix and the other as the minor helix. To mathematically describe a superhelix, we first construct the position vector of a regular axial helix (the major helix) to be $\H(\bs)= [A_M \cos(k_M \alpha \bs), A_M \sin(k_M \alpha \bs), \alpha \bs]$, with $\alpha = 1/\sqrt{1+A_M^2 k_M^2}$. Here $k_M$ is the wave-number, $A_M$ is the amplitude, and $\bs \in [0, L']$ and $L'$ are the arc-length parameter and the length of the major helix, respectively. From this basic helix the local unit tangent $\hat{\t}_A$, unit normal $\hat{\n}_A$, and unit binormal $\hat{\b}_A=\hat{\t}_A \times \hat{\n}_A$ vectors are determined, and we take them to define a local coordinate system upon which the minor helix is constructed \cite{jung07}. The position vector of the combined superhelical shape is then given by
\begin{align}
%\X(\bs) = [ x(\bs), y(\bs), z(\bs) ]=  \H(\bs) + A_m \cos(k_m \bs) \hat{\n}_A(\bs)\pm A_m \sin(k_m \bs) \hat{\b}_A(\bs), 
\X(\bs) = & \ [ x(\bs), y(\bs), z(\bs) ] \\
=  & \ \H(\bs) + A_m \cos(k_m \bs) \hat{\n}_A(\bs) \notag \\
&\pm A_m \sin(k_m \bs) \hat{\b}_A(\bs), 
\end{align}
where $A_m$ and $k_m$ are the amplitude and wave-number of the minor helix, respectively. Two different configurations will be considered: the `$+$' sign leads to a superhelical structure where the major and minor helices both have the same chirality, whereas the `$-$' sign represents the case of opposite chirality. Note that $\bs$ is no longer the natural arc-length parameter, but merely a regular parameter for describing the swimmer's geometry. The arc-length of the complete superhelix, denoted by $s$, as a function of the parameter $s'$ is determined by numerical integration, and the total length of the superhelix is denoted by $L$.

The major and minor helices are free to propagate waves at different wave speeds. Denoting $c_M$ and $c_m$ as the major and minor wave speeds respectively, the position vector at time $t$, $\X(\bs,t)$, may be written in component form as
\begin{align}
x (\bs,t) = & \ A_M \cos[k_M (\alpha \bs-c_M t)] \notag \\
& \ -A_m \cos[k_m (\bs-c_m t)]\cos[k_M (\alpha \bs-c_M t)] \notag \\
& \ \pm \alpha A_m \sin[k_m (\bs-c_m t)]\sin[k_M (\alpha \bs-c_M t)],\label{eqn:positionx} \\
y (\bs,t) = & \ A_M \sin[k_M (\alpha \bs-c_M t)] \notag \\
& \ -A_m \cos[k_m (\bs-c_m t)]\sin[k_M (\alpha  \bs-c_M t)] \notag \\
& \ \mp \alpha A_m \sin[k_m (\bs-c_m t)]\cos[k_M (\alpha  \bs-c_M t)], \label{eqn:positiony}\\
z (\bs,t) = & \ \pm \alpha A_m A_M k_M \sin[k_m (\bs-c_m t)]+\alpha \bs \label{eqn:positionz}.
\end{align}

\begin{figure*}
\begin{center}
\includegraphics[width=0.85\textwidth]{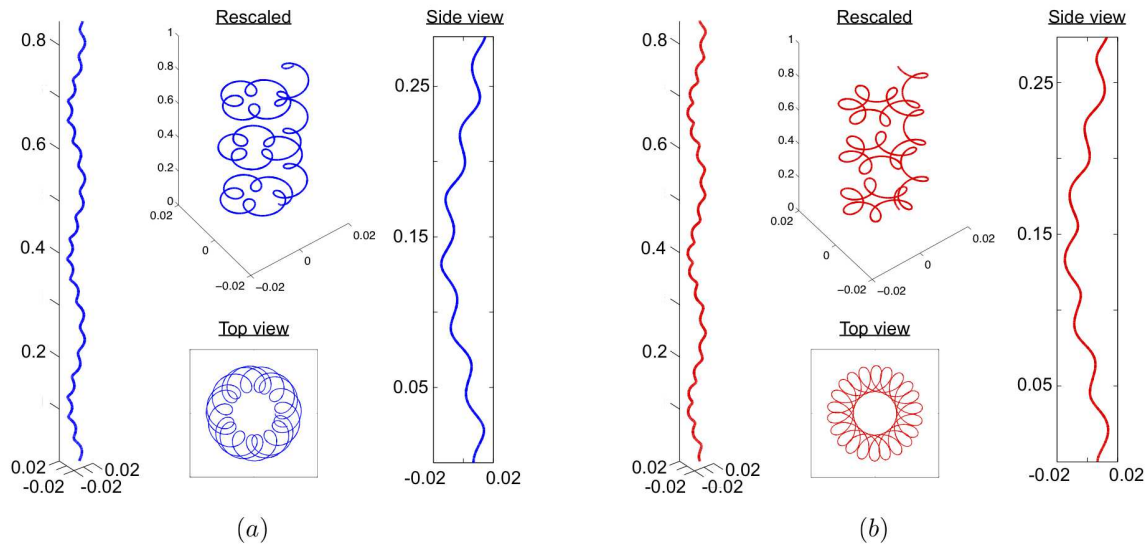}
\end{center}
\caption{\footnotesize  Idealization of the double-wave structure as superhelices for the (a) same-chirality, and (b) opposite-chirality configurations, using the dimensionless parameters of \textit{Culicoides melleus} spermatozoa (Table~\ref{table:dimensionless}).}
\label{fig:geo}
\end{figure*}

In addition to the long flagellum, insect spermatozoa have cell bodies which are very slender and short when compared to the flagellum size. We expect that the hydrodynamic influence of the sperm head is negligible and do not include such a body in our consideration. A typical superhelical swimmer is shown in Fig.~\ref{fig:geo}, using the dimensionless parameters of \textit{Culicoides melleus} spermatozoa.

\subsection{Hydrodynamical modeling}

Flagellar swimming is a result of the interaction between the actuating body and the surrounding fluid. A tractable and accurate approach to studying such hydrodynamic interactions exploits the slenderness of the flagellum, in which the velocity along the flagellar centerline is related to the fluid forces along the same curve. In previous studies, the fluid-body interaction has been modeled using a resistive force theory  \cite{wiggins98, yu06, kruse07}, in which local forces acting on the flagellum at any station along the filament are expressed in terms of the local velocity at the same location. Resistive force theory takes only local effects into account and neglects any hydrodynamic interactions between different parts of the deforming body. The local theory works well for simple geometries. However, due to the complexity of the superhelical structure in the problem under consideration here, the local theory is inadequate (see Section \ref{sec:SBTvsRFT} for details), and we employ instead the full non-local slender body theory \cite{johnson80} to study the hydrodynamics of the superhelical swimmers. The non-local theory captures the hydrodynamic interactions between distant parts of a curved filament, while still taking advantage of the slenderness of flagellum to simplify the analysis.

The flagellum is modeled as a slender filament of length $L$ and circular cross-section of radius $\epsilon\, L\, r(s)$, where $\epsilon \ll 1$ is the small aspect ratio of the flagellum (the maximum radius along the flagellum $a_r$ divided by its total length $L$), and $r(s)$ is the dimensionless radius. The non-local slender body theory is algebraically accurate in ratio of the slenderness $\epsilon$; namely, by setting the radius profile to $r(s) = \sqrt{4 s (L-s)}/L$, the computed fluid velocity is accurate to $O(\epsilon^2)$ (see \cite{johnson80}).

For a given velocity distribution $\v(s,t)$ along the filament at time $t$, the corresponding fluid force per unit length $\f(s,t)$ is given implicitly by the non-local relation
\begin{gather}
8 \pi \mu \v(s,t) = - \mathbf{\Lambda}[\f(s,t)] - \mathbf{K}[\f(s,t)], \label{eqn:SBT}
\end{gather}
where
\begin{align}
\mathbf{\Lambda}[\f](s,t)= & \left[c_0 (\I + \hat{\t} \hat{\t})+2(\I - \hat{\t} \hat{\t})\right] \f(s,t),\\
\mathbf{K}[\f](s,t)= & \ (\I+\hat{\t} \hat{\t}) \int_0^L \frac{\f(\tilde{s},t)-\f(s,t)}{|\tilde{s}-s|}\,d\tilde{s} \notag\\
& \ + \int_0^L \left(\frac{\I+\hat{\R} \hat{\R}}{|\R(\tilde{s},s; t)|}-\frac{\I+\hat{\t}\hat{\t}}{|\tilde{s}-s|}\right)\f(\tilde{s},t)\,d\tilde{s},
\end{align}
are the local and non-local operators, respectively, $\mu$ is the shear viscosity of the fluid, $c_0=-\ln(\epsilon^2 \text{e})>0$, $\R(\tilde{s},s; t)=\X(\tilde{s},t)-\X(s,t)$, $\hat{\R} = \R/|\R|$, and $\hat{\t}$ is the local unit tangent vector at the point $s$. 

The swimmers of interest deform their shapes in a prescribed, time-varying fashion (the superhelical wave pattern, Eqs.~\ref{eqn:positionx} to~\ref{eqn:positionz}), and the velocity created on its surface by this deformation is given by $\v_{\text{deform}} (s,t)= \partial \X(s,t)/ \partial t$. At every time instant $t$, the swimmer can be seen as a solid body with unknown translational velocity $\mathbf{U}(t)$ and rotation rate $\mathbf{\Omega}(t)$. The velocity created on the swimmer's surface due to swimming is then $\v_{\text{swim}}(s,t) = \U + \mathbf{\Omega} \times [\X(s,t)-\X_0]$, where $\X_0$ is an arbitrary reference point (taken as the origin here for simplicity). Therefore, the local velocity relative to the fluid $\v(s,t)$ is given by the sum of the deformation and swimming velocities: $\v(s,t)=\v_{\text{deform}}+\v_{\text{swim}} =  \U + \mathbf{\Omega} \times [\X(s,t)-\X_0]+\partial \X(s,t)/\partial t$. In this work, the wave propagation is towards the positive $z$ direction. Therefore, a negative swimming velocity ($\U$) means the propulsion occurs in a direction opposite to the wave propagation, while a positive swimming velocity means both the propulsion and wave propagation occur in the same direction.

Using a Galerkin method \cite{atkinson}, we express the local force $\f(s,t)$ as a finite sum of Legendre polynomials and solve Eq.~\ref{eqn:SBT} for $\f(s)$ by requiring the equation to hold under inner products against the same basis functions. The first integral in the non-local operator $\mathbf{K}[\f]$ is diagonalized in this space \cite{Gotzthesis,tornberg}. The system is closed by requiring the entire swimmer to be force free and torque free, 
\begin{gather}
\int_0^L \f(s) ds = 0, \label{eqn:forcebalance}\\
\int_0^L  \left[\X(s)-\X_0\right]\times \f(s) ds = 0, \label{eqn:torquebalance}
\end{gather}
providing at each moment in time a system of 6 equations to solve for the 6 unknowns $\mathbf{U}(t)$ and $\mathbf{\Omega}(t)$.

The swimming velocities determined in the manner described above represent velocities in a reference frame fixed on the swimmer. In order to study the full three-dimensional swimming kinematics in the laboratory frame (in which the body moves with velocity $\tilde{\mathbf{U}}(t)$ and rotation rate $\tilde{\mathbf{\Omega}}(t)$) we must include a transformation between the two. We denote the Cartesian coordinate system moving with the swimmer, the body frame, as $[\mathbf{e}_x,\mathbf{e}_y,\mathbf{e}_z]$ and the Cartesian coordinate system in the laboratory frame as $[\mathbf{e}_1,\mathbf{e}_2,\mathbf{e}_3]$.  
The evolution of the body frame, with respect to the laboratory frame, is then governed by
\begin{align}
\frac{d \mathbf{E}}{dt} =\mathbf{\Omega}(t)\times \mathbf{E}\label{eqn:frame},
\end{align}
where $\mathbf{E} = [\mathbf{e}_x,\mathbf{e}_y,\mathbf{e}_z] ^T$, along with the initial condition $[\mathbf{e}_x,\mathbf{e}_y,\mathbf{e}_z](t=0) = [\mathbf{e}_1,\mathbf{e}_2,\mathbf{e}_3]$.

\subsection{\label{sec:nondimensionalization}Non-dimensionalization}
The process of non-dimensionalization is very useful in science and engineering to identify the relevant dimensionless parameters governing the physics of the problem. In theoretical studies, it always allows a more concise description of the system, while in experimental studies, it reduces the number of independent experiments required to fully explore the problem. The present system is made dimensionless by scaling lengths by $1/k_m$, velocities by $c_m$, and time by their ratio, $1/c_m k_m$. The dimensionless position vector describing the kinematics of the superhelix is thus given by
\begin{align}
x = & \ R \bigl\{ \cos[K(\alpha \bs - c t)]-r \cos(\bs-t)\cos[K(\alpha \bs - c t)] \notag \\
&+\alpha r \sin(\bs-t)\sin[K(\alpha \bs - c\,t)]\bigr\}, \label{eqn:deformationX} \\
y = & \ R \bigl\{ \sin[K(\alpha \bs - c\,t)]-r \cos(\bs-t)\sin[K(\alpha \bs - c\,t)] \notag \\
&-\alpha r \sin(\bs-t)\cos[K(\alpha \bs - c\,t)] \bigr\}, \\
z =& \  \alpha \bs+ \alpha r R^2 K  \sin(\bs-t), \label{eqn:deformationZ}
\end{align}
where all variables are now understood to be dimensionless. Four dimensionless parameters characterizing the kinematics are identified above: the dimensionless amplitude of the major helix, $R = A_M k_m$, the ratio of the wave-numbers characterizing the major and minor helices, $K=k_M/k_m$, the ratio of the minor helix amplitude to the major helix amplitude, $r = A_m/A_M$, and the ratio of the major wave speed to the minor wave speed $c=c_M/c_m$. In the double-wave pattern observed in insect spermatozoa, the major wave amplitude is always larger than the minor wave amplitude ($r<1$; though this need not be true for a general superhelix). In addition, it is also observed that the minor wave speed is always greater than the major wave speed in the double-wave structure of insect spermatozoa ($c<1$).

\subsection{\label{sec:data}Kinematic and geometric data}
From the non-dimensionalization above, we have identified four dimensionless parameters ($r, R, K, c$) required to fully characterize the centerline motion of a superhelical flagellum. Werner and Simmons \cite{werner08} have compiled a very useful table containing kinematic and geometric data of the double-wave structure observed in insect spermatozoa in previous studies. The table reveals that experimental measurements of the necessary quantities for hydrodynamic modeling are very limited, due to the difficulties involved in interpreting three-dimensional data from two-dimensional images \cite{linley79}.

Here we follow the table compiled by Werner and Simmons and estimate the missing information based on images of insect spermatozoa reported in the literature. Table~\ref{table:geometry} contains reported and estimated data on the double-wave structure, including the wavelengths, amplitudes, frequencies, major and minor wave speeds, flagellum thickness, and the swimming speed of different insect spermatozoa. Our estimated quantities are marked with stars to distinguish them from reported quantities by the original papers. In some studies (e.g.~for \textit{Lygaeus} \cite{philips74}), only relative lengths can be given due to the lack of scale bars in the reported images; these relative quantities are square-bracketed in Tables~\ref{table:geometry} \&~\ref{table:dimensionless}. Geometric and kinematic data can display large variations even within a species (\textit{Tenebrio molitor} \cite{baccetti73a}). When the distribution of the quantities are not given, arithmetic means of the available measurements are used whenever appropriate in the present study.

The measurements are presented in the corresponding dimensionless quantities in Table~\ref{table:dimensionless}, using the scalings defined in \S\ref{sec:nondimensionalization}. We have not found in the referenced literature any information about the chirality of the major and minor helical structures. We therefore present results below for both the same-chirality and opposite-chirality configurations. In our measurements (collected in Table~\ref{table:geometry}), the wavelength of the minor wave is taken to be the two-dimensional distance between adjacent minor wave peaks. A small correction factor is required to convert these two-dimensional quantities to the appropriate wavelengths in describing the three-dimensional superhelical structure. The correction factor depends on whether the superhelical structure is in the same-chirality ($1/(1-\alpha \lambda_m/\lambda_M)$) or the opposite-chirality ($1/(1+\alpha \lambda_m/\lambda_M)$) configuration.

%

%%%%%%%%%% Tables
\begin{sidewaystable}[!] 
\centering       % centering table 
\caption{\label{table:geometry} \footnotesize  Kinematic and geometric data of the double-wave structure. We follow the table by Werner and Simmons \cite{werner08} with our estimations (marked with stars *) of the missing quantities based on the images of the sperm cell reported in the references. $a_r$ refers to the radius of the flagellum.}  % title name of the table 
\centering       % centering table 
\setlength{\tabcolsep}{1pt} 
\begin{tabular}{l c c c c c c c c c c c c c c} \hline \hline
% the top line of the headings 
\multicolumn{1}{c}{} & 
\multicolumn{4}{c}{Major Wave} &
\multicolumn{1}{c}{} &  %for spacing
\multicolumn{4}{c}{Minor Wave} &
\multicolumn{1}{c}{} &  %for spacing
\multicolumn{1}{c}{$2a_r$} &
\multicolumn{1}{c}{} &  %for spacing
\multicolumn{1}{c}{$V_{\text{sperm}}$} &
\multicolumn{1}{c}{References} \\ \cline{2-5} \cline{7-10} % for the head size and the sperm speed and remark

% the middle line of the headings 
\multicolumn{1}{l}{Species} & 
\multicolumn{1}{c}{$\lambda_M \ $($\mu$m)} &
\multicolumn{1}{c}{$A_M \ $($\mu$m)} &
\multicolumn{1}{c}{$f_M \ $(Hz)} &
\multicolumn{1}{c}{$c_M \ $ ($\mu$m/s)} &
\multicolumn{1}{c}{} &  %for spacing
\multicolumn{1}{c}{$\lambda_s \ $($\mu$m)} &
\multicolumn{1}{c}{$A_m \ $($\mu$m)} &
\multicolumn{1}{c}{$f_m \ $(Hz)} &
\multicolumn{1}{c}{$c_m \ $ ($\mu$m/s)} &
\multicolumn{1}{c}{} &  %for spacing
\multicolumn{1}{c}{($\mu$m)} &
\multicolumn{1}{c}{} &  %for spacing
\multicolumn{1}{c}{($\mu$m)}&
\multicolumn{1}{l}{}\\ [0.5 ex]  \hline

\multicolumn{1}{l}{\textit{Aedes notoscriptus}}&
\multicolumn{1}{c}{$28^*$}&
\multicolumn{1}{c}{$3.67^*$}&
\multicolumn{1}{c}{$3.4$}&
\multicolumn{1}{c}{$95.2^*$}&
\multicolumn{1}{c}{$$}&
\multicolumn{1}{c}{$6.8^*$}&
\multicolumn{1}{c}{$1.83^*$}&
\multicolumn{1}{c}{$34$}&
\multicolumn{1}{c}{$231.2^*$}&
\multicolumn{1}{c}{}&
\multicolumn{1}{c}{$0.55^*$}&
\multicolumn{1}{c}{}&
\multicolumn{1}{c}{} &  %for spacing
\multicolumn{1}{c}{\cite{swan81}}\\

\multicolumn{1}{l}{\textit{Bacillus rossius}}&
\multicolumn{1}{c}{$40^*$}&
\multicolumn{1}{c}{$12$}&
\multicolumn{1}{c}{12}&
\multicolumn{1}{c}{$480^*$}&
\multicolumn{1}{c}{}&
\multicolumn{1}{c}{$17.9^*$}&
\multicolumn{1}{c}{1.1}&
\multicolumn{1}{c}{40}&
\multicolumn{1}{c}{$716^*$}&
\multicolumn{1}{c}{}&
\multicolumn{1}{c}{$0.74^*$}&
\multicolumn{1}{c}{}&
\multicolumn{1}{c}{} &  %for spacing
\multicolumn{1}{c}{\cite{baccetti73b}}\\

\multicolumn{1}{l}{\textit{Ceratitis capitata}\footnote[1]{Only forward mode is considered (see \cite{baccetti89}).}}&
\multicolumn{1}{c}{30}&
\multicolumn{1}{c}{20}&
\multicolumn{1}{c}{4}& 
\multicolumn{1}{c}{$120^*$}&
\multicolumn{1}{c}{}&
\multicolumn{1}{c}{7-8}&
\multicolumn{1}{c}{1-2}&
\multicolumn{1}{c}{20}&
\multicolumn{1}{c}{$150^*$}&
\multicolumn{1}{c}{}&
\multicolumn{1}{c}{$0.68^*$}&
\multicolumn{1}{c}{} &  %for spacing
\multicolumn{1}{c}{16}&
\multicolumn{1}{c}{\cite{baccetti89}}\\

\multicolumn{1}{l}{\textit{Culicoides melleus}}&
\multicolumn{1}{c}{54.1}&
\multicolumn{1}{c}{2.1}&
\multicolumn{1}{c}{}&
\multicolumn{1}{c}{}&
\multicolumn{1}{c}{}&
\multicolumn{1}{c}{8.7}&
\multicolumn{1}{c}{0.8}&
\multicolumn{1}{c}{8.2}&
\multicolumn{1}{c}{80}&
\multicolumn{1}{c}{}&
\multicolumn{1}{c}{$0.77^*$}&
\multicolumn{1}{c}{} &  %for spacing
\multicolumn{1}{c}{8.3}&
\multicolumn{1}{c}{\cite{linley79}}\\

\multicolumn{1}{l}{\textit{Drosophila obscura}\footnote[2]{Only male long sperm is considered here (see \cite{bressac91}).}}&
\multicolumn{1}{c}{$45^*$}&
\multicolumn{1}{c}{16}&
\multicolumn{1}{c}{}&
\multicolumn{1}{c}{}&
\multicolumn{1}{c}{}&
\multicolumn{1}{c}{$13.3^*$}&
\multicolumn{1}{c}{$0.5^*$}&
\multicolumn{1}{c}{20.4}&
\multicolumn{1}{c}{271}&
\multicolumn{1}{c}{}&
\multicolumn{1}{c}{$0.76^*$}&
\multicolumn{1}{c}{}&
\multicolumn{1}{c}{} &  %for spacing
\multicolumn{1}{c}{\cite{bressac91}}\\

\multicolumn{1}{l}{\textit{Lygaeus}\footnote[3]{Scale bars were not presented with the images in the reference and only relative lengths (square-bracketed) can be estimated.}}&
\multicolumn{1}{c}{$[1]^*$}&
\multicolumn{1}{c}{$[0.1]^*$}&
\multicolumn{1}{c}{}&
\multicolumn{1}{c}{}&
\multicolumn{1}{c}{}&
\multicolumn{1}{c}{$[0.21]^*$}&
\multicolumn{1}{c}{$[0.014]^*$}&
\multicolumn{1}{c}{130}&
\multicolumn{1}{c}{}&
\multicolumn{1}{c}{}&
\multicolumn{1}{c}{$[0.024]^*$}&
\multicolumn{1}{c}{}&
\multicolumn{1}{c}{} &  %for spacing
\multicolumn{1}{c}{\cite{philips74}}\\

\multicolumn{1}{l}{\textit{Megaselia scalaris}}&
\multicolumn{1}{c}{$68^*$}&
\multicolumn{1}{c}{$9.3^*$}&
\multicolumn{1}{c}{3.1}&
\multicolumn{1}{c}{$210.8^*$}&
\multicolumn{1}{c}{}&
\multicolumn{1}{c}{$7^*$}&
\multicolumn{1}{c}{$0.5^*$}&
\multicolumn{1}{c}{}&
\multicolumn{1}{c}{}&
\multicolumn{1}{c}{}&
\multicolumn{1}{c}{$0.78^*$}&
\multicolumn{1}{c}{} &  %for spacing
\multicolumn{1}{c}{12.7}&
\multicolumn{1}{c}{\cite{curtis91}}\\

\multicolumn{1}{l}{\textit{Tenebrio molitor}\footnote[4]{Round-brackets denote arithmetic means.}}&
\multicolumn{1}{c}{20-30(25)}& 
\multicolumn{1}{c}{9-15(12)}&
\multicolumn{1}{c}{0.9-2.8(1.85)}&
\multicolumn{1}{c}{20-90(55)}&
\multicolumn{1}{c}{}&
\multicolumn{1}{c}{6-12(9)}&
\multicolumn{1}{c}{3-4(3.5)}&
\multicolumn{1}{c}{7-28(17.5)}&
\multicolumn{1}{c}{40-300(170)}&
\multicolumn{1}{c}{}&
\multicolumn{1}{c}{$0.79^*$}&
\multicolumn{1}{c}{} &  %for spacing
\multicolumn{1}{c}{16-100(58)}&
\multicolumn{1}{c}{\cite{baccetti73a}}\\
[0.5 ex]\hline \hline
\end{tabular}
\label{tab:LPer} 
\end{sidewaystable} 
%%%%%%%%%%%%%%%%%%%%%%%%%%%%%%%%%%%%%%

\begin{sidewaystable}[!]
\caption{\label{table:dimensionless} \footnotesize  Dimensionless parameters of superhelical flagella: $K=k_M/k_m; \ R=A_M k_m; \ r = A_m/A_M; \ c= c_M/c_m; \ V = V_{\text{sperm}}/c_m; \ a=a_{r}k_m$; $N$ refers to number of large wavelengths. $\tilde{U}_+$ and $\tilde{U}_-$ are the predicted average swimming velocities for the same-chirality and opposite-chirality configurations respectively.}  % title name of the table 
\centering       % centering table 
\setlength{\tabcolsep}{8pt} 
\begin{tabular}{l c c c c c c c c c c} \hline \hline
% the top line of the headings 
\multicolumn{1}{c}{Species} & 
\multicolumn{1}{c}{$K$} &
\multicolumn{1}{c}{$R$} &
\multicolumn{1}{c}{$r$} &
\multicolumn{1}{c}{$c$} &
\multicolumn{1}{l}{$N$} &
\multicolumn{1}{l}{$a$} &
\multicolumn{1}{c}{$V$} &
\multicolumn{1}{c}{$\tilde{U}_+ \ (\tilde{U}_-)$} &
\multicolumn{1}{c}{$\eta_+ \ (\eta_-)$}&
\multicolumn{1}{c}{$|\tilde{U}_+-V|/V \ (|\tilde{U}_- -V|/V)$} 
\\ \cline{1-11}

% the middle line of the headings 

\multicolumn{1}{l}{\textit{Aedes notoscriptus}}&
\multicolumn{1}{c}{$0.24^*$}&
\multicolumn{1}{c}{$3.4^*$}&
\multicolumn{1}{c}{$0.5^*$}&
\multicolumn{1}{c}{$0.41^*$}&
\multicolumn{1}{c}{1}&
\multicolumn{1}{c}{$0.25^*$}&
\multicolumn{1}{c}{}&
\multicolumn{1}{c}{0.21 (0.16)}&
\multicolumn{1}{c}{1.2\%(0.34\%)}&
\multicolumn{1}{c}{}
\\

\multicolumn{1}{l}{\textit{Bacillus rossius}}&
\multicolumn{1}{c}{$0.45^*$}&
\multicolumn{1}{c}{$4.2^*$}&
\multicolumn{1}{c}{0.092}&
\multicolumn{1}{c}{$0.67^*$}&
\multicolumn{1}{c}{1.5}&
\multicolumn{1}{c}{$0.13^*$}&
\multicolumn{1}{c}{}&
\multicolumn{1}{c}{0.031 (0.033)}&
\multicolumn{1}{c}{0.30\%(0.14\%)}&
\multicolumn{1}{c}{}

\\

\multicolumn{1}{l}{\textit{Ceratitis capitata}}&
\multicolumn{1}{c}{0.25}&
\multicolumn{1}{c}{16.8}&
\multicolumn{1}{c}{0.075}&
\multicolumn{1}{c}{$0.8^*$}&
\multicolumn{1}{c}{$4-5$\footnote[5]{A value of 4 is adopted in the simulation.}}&
\multicolumn{1}{c}{$0.28^*$}&
\multicolumn{1}{c}{$0.11^*$}&
\multicolumn{1}{c}{0.15 (0.039\footnote[6]{velocity in the $z$-direction occurs in the same direction as the wave propagation.})}&
\multicolumn{1}{c}{0.40\% (0.021\%)}&
\multicolumn{1}{c}{36\% (65\%)}
\\

\multicolumn{1}{l}{\textit{Culicoides melleus}}&
\multicolumn{1}{c}{0.16}&
\multicolumn{1}{c}{1.5}&
\multicolumn{1}{c}{0.38}&
\multicolumn{1}{c}{}&
\multicolumn{1}{c}{3}&
\multicolumn{1}{c}{$0.28^*$}&
\multicolumn{1}{c}{0.10}&
\multicolumn{1}{c}{0.051 (0.15)}&
\multicolumn{1}{c}{0.30\% (1.5\%)}&
\multicolumn{1}{c}{49\% (50\%)}
\\

\multicolumn{1}{l}{\textit{Drosophila obscura}}&
\multicolumn{1}{c}{$0.30^*$}&
\multicolumn{1}{c}{$7.6^*$}&
\multicolumn{1}{c}{$0.03^*$}&
\multicolumn{1}{c}{}&
\multicolumn{1}{c}{1}&
\multicolumn{1}{c}{$0.18^*$}&
\multicolumn{1}{c}{}&
\multicolumn{1}{c}{0.012 (0.0096)}&
\multicolumn{1}{c}{0.10\%(0.036\%)}&
\multicolumn{1}{c}{}
\\

\multicolumn{1}{l}{\textit{Lygaeus}}&
\multicolumn{1}{c}{$0.21^*$}&
\multicolumn{1}{c}{$3^*$}&
\multicolumn{1}{c}{$0.14^*$}&
\multicolumn{1}{c}{}&
\multicolumn{1}{c}{$1-2$\footnote[7]{A value of 1 is adopted in the simulation.}}&
\multicolumn{1}{c}{$0.34^*$}&
\multicolumn{1}{c}{}&
\multicolumn{1}{c}{0.033 (0.081)}&
\multicolumn{1}{c}{0.24\%(0.71\%)}&
\multicolumn{1}{c}{}
\\

\multicolumn{1}{l}{\textit{Megaselia scalaris}}&
\multicolumn{1}{c}{$0.10^*$}&
\multicolumn{1}{c}{$8.3^*$}&
\multicolumn{1}{c}{$0.05^*$}&
\multicolumn{1}{c}{}&
\multicolumn{1}{c}{1}&
\multicolumn{1}{c}{$0.35^*$}&
\multicolumn{1}{c}{}&
\multicolumn{1}{c}{0.035 (0.065)}&
\multicolumn{1}{c}{0.16\%(0.42\%)}&
\multicolumn{1}{c}{}
\\

\multicolumn{1}{l}{\textit{Tenebrio molitor}}&
\multicolumn{1}{c}{0.36}&
\multicolumn{1}{c}{8.4}&
\multicolumn{1}{c}{0.29}&
\multicolumn{1}{c}{0.32}&
\multicolumn{1}{c}{4}&
\multicolumn{1}{c}{$0.28^*$}&
\multicolumn{1}{c}{0.34}&
\multicolumn{1}{c}{0.40 (0.41\footnote[8]{velocity in the $z$-direction occurs in the same direction as the wave propagation.})}&
\multicolumn{1}{c}{1.2\%(0.86\%)}&
\multicolumn{1}{c}{18\% (21\%)}
\\
[0.5 ex]\hline \hline
\end{tabular}
\end{sidewaystable}

\section{\label{sec:results}Results}
In this section, we first use the framework described above to predict the swimming performance of spermatozoa of different species, and compare our hydrodynamic results with available experimental measurements. We then illustrate the basic features of superhelical swimming by focusing on a model organism, namely \textit{Culicoides melleus}, and proceed to perform a parametric study investigating the effects of certain kinematic parameters (minor and major wave speeds) and geometric parameters (minor and major wave amplitudes). 

 \subsection{\label{sec:performance}Hydrodynamic performance}
 
\subsubsection{Propulsion speed}
The propulsion speed is an important functional parameter characterizing the motility of a sperm cell. There exist only very few measurements of the swimming speed of insect spermatozoa exhibiting the double-wave structure: \textit{Ceratitis capitata} was observed to swim with speed $V_{\text{sperm}}=16$ $\mu$m/s \cite{baccetti89} and the minor wave speed is estimated to be 150 $\mu$m/s (hence, $V=V_{\text{sperm}}/c_m$=0.11); \textit{Culicoides melleus} \cite{linley79} was observed to swim with speed 8.3 $\mu$m/s and minor wave speed 80 $\mu$m/s ($V=0.10$). For the case of \textit{Tenebrio molitor}, a wide range of sperm speeds were reported (from 16 $\mu$m/s to 100 $\mu$m/s) as well as minor wave speeds (40 $\mu$m/s to 300 $\mu$m/s) \cite{baccetti73a}. The distributions were not reported however, so using the arithmetic means we obtain a swimming speed to minor wave speed ratio of $V=0.34$.

\begin{figure*}
\begin{center}
\includegraphics[width=0.85\textwidth]{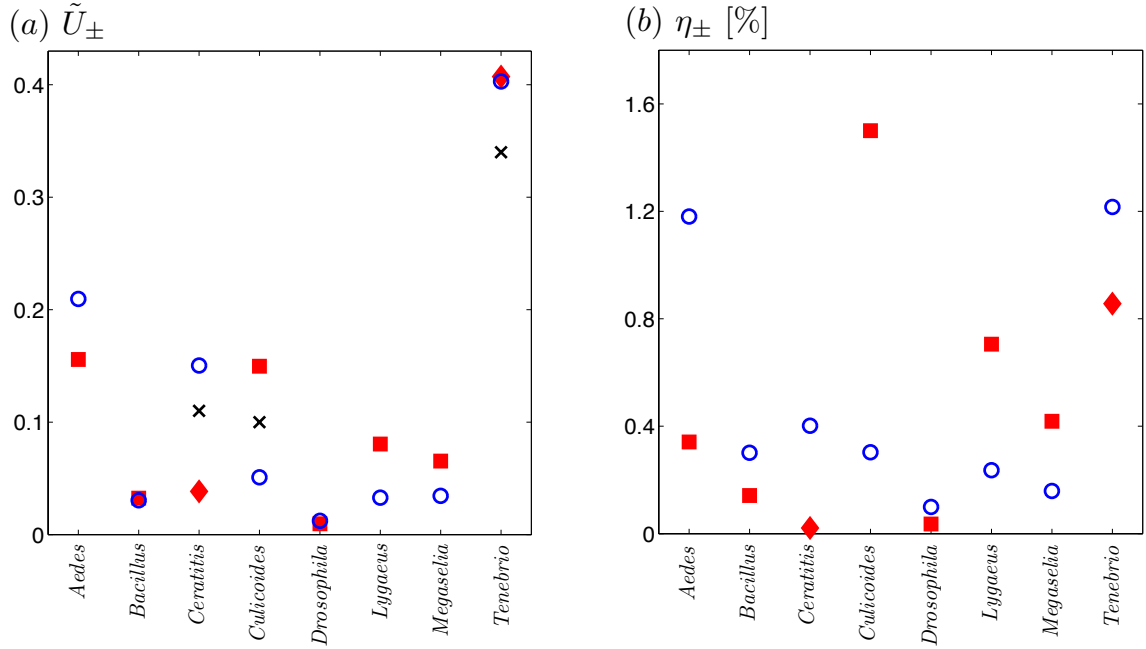}
\end{center}
\caption{\footnotesize  Predicted swimming performance: (a) average swimming speed, $\tilde{U}_{\pm}$, of different species; (b) hydrodynamic efficiency, $\eta_{\pm}$, of different species. Open symbols (blue circles) represent the same-chirality configuration ($+$); filled symbols (red squares and diamonds) represent the opposite-chirality configuration ($-$); for blue open circles and red squares, swimming occurs in the opposite direction as the wave propagation; on the contrary, red diamonds represent the cases of opposite-chirality configuration where the velocity in the $z$-direction occurring in the same direction as the wave propagation; black crosses represent experimental measurements of $V = V_{\text{sperm}}/c_m$ (See Tables \ref{table:geometry} and \ref{table:dimensionless}).}
\label{fig:performance}
\end{figure*}

For a superhelical swimmer in our model, the swimming kinematics are three-dimensional and unsteady in time \footnote[2]{The deformations (Eqs.~\ref{eqn:positionx} to \ref{eqn:positionz}) are higher frequency oscillations modulated by lower frequency oscillation (a beat). There does not exists a time period $T$ such that the deformation vector repeats itself: $\R(s,t+T) \neq \R(s,t),  \forall T$.}, and the most relevant quantification of the propulsion speed is an average swimming velocity in the laboratory frame, denoted by $\tilde{U}_{\pm}$, where the {$+$} and $-$ signs represent the same- and opposite-chirality configurations respectively. The average swimming speed is defined as $\tilde{U}_{\pm}=|[ \langle \tilde{U}_x \rangle, \langle \tilde{U}_y \rangle,\langle \tilde{U}_z \rangle  ]|$, where $|...|$ denotes the magnitude of a vector, and $\langle ... \rangle$ denotes a time average \footnote{The time averaging is defined for a function $f(t)$ as $\int^{t_0}_0 f(t) dt/t_0$, where $t_0$ is a sufficiently large time such that the peak-to-peak fluctuation in time is less than 1\% of the final average value}.  Since the chirality configuration remains unknown, we present predictions for both cases in Fig.~\ref{fig:performance}a. Only the three sets of experimental measurements of the swimming performance ($V=V_{\text{sperm}}/c_m$) mentioned above are available for comparison and are super-imposed on the same figure (see Fig.~\ref{fig:performance} caption). Predictions as ratios of the swimming speed relative to the minor wave speed (dimensionless speed) for species with no measurement of $V$ are also provided in Fig.~\ref{fig:performance}a.

In most of the cases considered, the propulsion in the longitudinal direction ($z$-direction) of these superhelical swimmers is opposite to the direction of wave propagation (the major wave propagates in the positive $z$-direction, and the minor wave propagates along the curved major helix and distally towards the positive $z$-direction), regardless of the chirality configuration. This is not unlike the behavior of swimmers propagating a planar sinusoidal or a regular helical wave, for which the swimming direction is also opposite the direction of the wave propagation. In superhelical swimming, however, we also find cases of the opposite-chirality configuration where the body swims in the same direction as the wave propagation. Specifically, we note a qualitative difference between the study of the same-chirality and opposite-chirality configurations for \textit{Ceratitis capitata} and \textit{Tenebrio molitor}. In these simulations, a superhelical wave is set to propagate in the positive $z$-direction, and both the opposite-chirality configurations of \textit{Ceratitis capitata} and \textit{Tenebrio molitor} generate a positive $U_z$ (swimming is in the same direction as the wave propagation), while their corresponding same-chirality configurations generate a negative $U_z$ (swimming is in the opposite direction as the wave propagation). The speed and efficiency of these peculiar swimmers are distinguished from other cases by red diamonds in Fig.~\ref{fig:performance}. It remains a question whether this phenomenon may be observed in nature, since the chirality configuration and the swimming direction (relative to the wave propagation) of actual insect spermatozoa displaying the double-wave structure are still unclear. Nevertheless, this phenomenon by itself is intriguing and will be further explored in \S\ref{sec:c}.

The swimming speed predictions lie at least within the same order of magnitude of the experimental measurements for both chirality configurations. The same-chiralty results provide slightly better agreement than the opposite-chiralty results (without taking the swimming direction into account). For the same-chirality configuration, the discrepancies, $|\tilde{U}_{+} -V|/V$, between the predictions and the experimental measurements read 36\% for \textit{Ceratitis capitata}, 49\% for \textit{Culicoides melleus}, and 18\% for \textit{Tenebrio molitor}. For the opposite-chirality configuration, the discrepancy between our predictions and the experimental measurements, $|\tilde{U}_{-} -V|/V$, are 65\% for \textit{Ceratitis capitata}, 50\% for \textit{Culicoides melleus}, and 21\% for \textit{Tenebrio molitor}. Given the primitive nature of the data employed (see \S\ref{sec:data}), we consider the agreements here to be reasonable. However, we cannot draw definite conclusions on the issue of chirality configuration; further experimental observations are necessary.  

It shall be remarked that the present study is largely constrained by the unavailability of experimental measurement data. Critical kinematic information, such as the major and minor wave speeds, are often not reported in the literature and are impossible to estimate from the images available. The speed ratio $c=c_M/c_m$ is not available for simulations for most species. The computations here are still possible because of the independence of the swimming kinematics on the parameter $c$ (verified numerically and will be explained in \S\ref{sec:c}). Therefore, the specific value of $c$ is unimportant for all cases considered here; we adopt $c=0$ in all simulations hereafter unless otherwise stated.

\subsubsection{\label{sec:efficiency}Hydrodynamic efficiency}
Another important functional parameter is the hydrodynamic efficiency of the swimmer. In the microscopic world, the hydrodynamic efficiency is typically very low. For a rigid helix, Lighthill \cite{lighthill75} calculated theoretically the maximum efficiency attainable to be about 8.5\%, while the typical efficiency of biological cells is around 1-2\% \cite{chattopadhyay06,childress, lighthill75,lauga2,sl10,sl11}. For low Reynolds number swimming, a common measure of the hydrodynamic efficiency, $\eta$, is the ratio of the rate of work required to drag the straightened flagellum through the fluid to the rate of work done on the fluid by the flagellum during swimming \cite{lighthill75},
\begin{align}
\eta_{\pm} = \frac{\xi_{\parallel} L \tilde{U}_{\pm}^2}{\left\langle \int_0^L \v \cdot \f ds \right\rangle},
\end{align}
where $\xi_{\parallel}= 4\pi/c_0$ is the drag coefficient for a straight, slender rod, and the brackets indicate a time average. We calculate the hydrodynamic efficiencies for both the same- and opposite-chirality configurations (see Fig.~\ref{fig:performance}b) for spermatozoa of different species. For the same-chirality configuration, the efficiency ranges from 0.16\% (\textit{Megaselia  scalaris}) to 1.2\% (\textit{Aedes notoscriptus} and \textit{Tenebrio molitor}); for the opposite-chirality configuration, the efficiency ranges from 0.036\% (\textit{Drosophila obscura}) to 1.5\% (\textit{Culicoides melleus}). The efficiencies of these swimmers are comparable to typical biological cells.

 \subsection{\label{sec:modelproblem}Model organism: \textit{Culicoides melleus}}
In this section, we illustrate the features of superhelical swimming by singling out a superhelical swimmer defined using the geometric data of the sperm cell of \textit{Culicoides melleus} (Table \ref{table:dimensionless}). See Fig.~\ref{fig:geo} for the swimmer geometry.
 
\subsubsection{Swimming kinematics}

\begin{figure*}
\begin{center}
\includegraphics[width=0.95\textwidth]{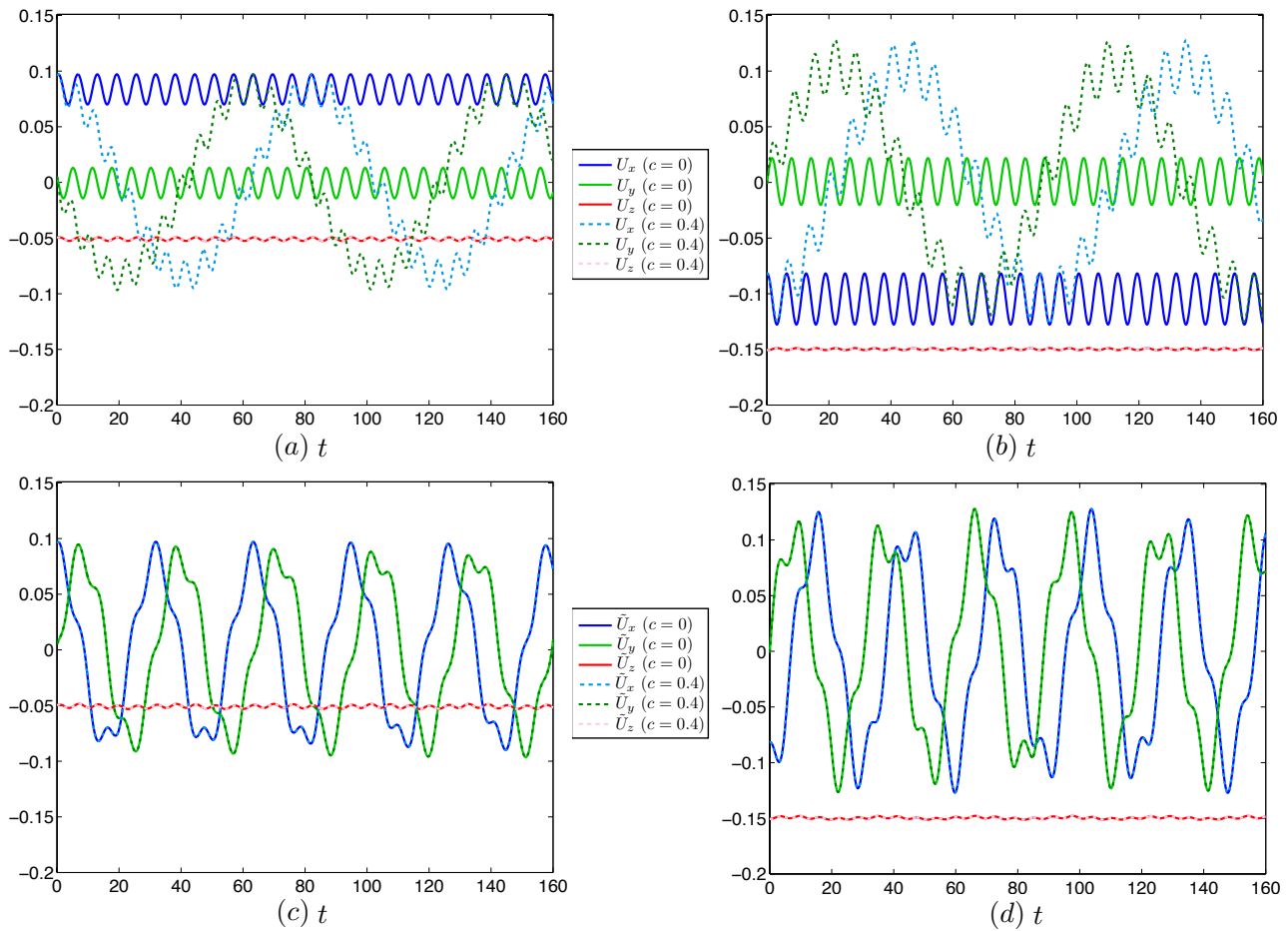}
\end{center}
\caption{\footnotesize  Three-dimensional swimming velocities in the body frame $\mathbf{U}= ( U_x, U_y, U_z )$ for (a) the same-chirality and (b) the opposite-chirality configurations, and in the laboratory frame $\tilde{\mathbf{U}}= ( \tilde{U}_x, \tilde{U}_y, \tilde{U}_z )$ for (c) the same-chirality and (d) the opposite-chirality configurations, for \textit{Culicoides melleus} spermatozoa.}
\label{fig:kinematics}
\end{figure*}

The swimming velocities computed for motion in the body frame $[U_x, U_y, U_z]$ and in the laboratory frame $[ \tilde{U}_x,\tilde{U}_y, \tilde{U}_z]$ of the superhelical swimmer with data from the sperm cell of \textit{Culicoides melleus} are plotted in Fig.~\ref{fig:kinematics} for the same-chirality (Fig.~\ref{fig:kinematics}a) and opposite-chirality (Fig.~\ref{fig:kinematics}b) configurations. We include in these figures the results for ratio of the major and minor wave speeds $c=0$ (solid lines) and $c=0.4$ (dotted lines). When $c=0$, the deformation (Eqs.~\ref{eqn:deformationX} to~\ref{eqn:deformationZ}) is $2\pi$-periodic, therefore the swimming kinematics in the body frame are also $2\pi$-periodic. However, when observed in the laboratory frame, the coupling between the translational and rotational kinematics renders the swimming velocities no longer 2$\pi$-periodic, and the motion is unsteady in time (Fig.~\ref{fig:kinematics}c \& d). We observe a pattern consisting of higher frequency oscillations modulated by a lower frequency envelope. For the case of $c=0.4$, when observed in the body frame, we see modulated waveforms. However, when transformed to the laboratory frame, the cases of $c=0$ and $c=0.4$ have identical swimming kinematics (see the overlapped solid and dotted lines in Fig.~\ref{fig:kinematics}c \& d), implying that the swimming kinematics in the laboratory frame are independent of the major propagating wave speed. The value of $c$ affects only the kinematics in the body frame; why the relative wavespeed is unimportant is described in greater detail in \S\ref{sec:c}.

The three-dimensional swimming velocities give rise to a doubly-helicated trajectory (the presence of a minor structure on top of a major helical structure). However, this could be difficult to observe experimentally, since the major amplitude of the doubly-helicated trajectory is usually much smaller than that of the superhelical swimmer; the swimmer would apparently move with a straight trajectory (with very small oscillations in the transverse direction). Recall that for regular helical swimming ($r=0$), the trajectory of the helical swimmer reduces to a regular helix.

\subsubsection{Head-less swimming}
The shape and size of the sperm head vary among spermatozoa of different species: human and bull spermatozoa have relatively large, paddle-shaped heads, whereas insect spermatozoa have elongated heads which are almost indistinguishable from the mid-piece. The additional hydrodynamic resistance from the presence of a head would seemingly degrade the swimming performance of the sperm cell. However, Chwang and Wu \cite{chwang71} showed that a sperm head is actually necessary for helical swimming; without one, the motion is that of a rotating rigid body, which cannot be realized absent an external force or torque \cite{keller76b}. To satisfy the zero net-torque condition, a sperm head is required to balance the reaction torque acting on the flagellum. This constraint does not apply for planar, sinusoidal wave motion, which can swim without an anchor or load. 

We pause to point out a subtle but important difference between a rotating prokaryotic helical tail and a eukaryotic tail propagating a bending helical wave. For a eukaryotic tail propagating a bending wave, the fluid forces act to rotate the flagellum opposite the direction of the apparent helical rotation. The rotation due to the fluid reaction creates torques due to local spinning (rotation of the flagellum about its centerline), which balance the opposing torque generated by the helical wave propagation. Therefore, a eukaryotic cell could theoretically swim without a sperm head, albeit very slowly because the flagellum is very slender and the local torques are correspondingly small. Quantitatively, using a local drag model (\cite{chwang71}) it can be shown that the head-less swimming speed scales as $U/c \sim 2 \mu k^2/\xi_{\parallel} (1+k^2 A^2) b^2 +O(b^4)$, where $b$ is the radius of a cross-section of the flagellum, and $k$ and $A$ are the wavenumber and helical radius respectively. Using the geometrical data of flagella of \textit{Euglena viridis} summarized by Brennen and Winet \cite{brennen}, and assuming a flagellar diameter of $2b\approx 0.25 \mu$m \cite{margulis, nicastro}, we find $U/c \approx 10^{-3}$. For a prokaryotic tail, however, the helix rotates as a rigid body. In this case, the local spinning torques generated by the active helical rotation and the passive rotation due to the fluid reaction are identical in magnitude but opposite in sign. The torque-free condition therefore requires that the fluid reaction counter-rotates the helix at precisely the rotation rate of the helical wave propagation. Hence there can be no effective helical rotation, and the body cannot swim. Although head-less swimming is theoretically possible for eukaryotic tails, the swimming speed would be exceptionally small as shown by the estimation above. The subtle difference between the two types of helical waves just described is often therefore neglected, and it is generally reasonable to state that head-less swimming is not possible using helical wave propagation.

In superhelical swimming a sperm head is not required for self-propulsion (indeed, all cases reported in this paper were studied with the absence of a sperm head) since a superhelical wave motion is in general not a rigid body motion. Furthermore, since actual insect spermatozoa heads are very slender and short compared with the entire length of the flagellum, the contribution of its hydrodynamic resistance and the hydrodynamic interactions with the flagellum should be negligible. Therefore, we do not expect the presence of a slender and short sperm head to introduce qualitative differences in the results.

\subsection{\label{sec:parametric}Parametric study}
Next, we explore the effects of certain kinematic (wave speeds) and geometric parameters (wave amplitudes) on the swimming velocities of a superhelical swimmer.

\subsubsection{\label{sec:c}The effect of $c$} 
In this problem, we have scaled the velocities upon the minor wave speed $c_m$, which implies that the dimensionless swimming velocities scale linearly with the minor wave speed. Here we examine the effect of the major wave speed on the swimming performance of a superhelical swimmer and answer the question:  to what relative extent do the two waves contribute to the propulsion of the superhelical swimmer \cite{linley79}? Specifically, we study the effect of the parameter $c = c_M / c_m$, which is the ratio of the major wave speed to the minor wave speed. For every species shown in Table~\ref{table:dimensionless}, we fix all parameters but vary the value of $c$ from zero to unity ($c <1$, as $c_M < c_m$). It is found that the swimming velocities in the laboratory frame is independent of the value of $c$. We have already shown in Fig.~\ref{fig:kinematics} the results for two values of $c$ for illustration. While the resulting swimming motions in these two cases differ significantly in the body frame, they are identical when observed in a laboratory frame of reference. The numerical results imply that the major wave speed does not contribution to propulsion.

These findings may be understood by noting an ambiguity in definition. First consider a single helical filament (without a sperm head) placed in a fluid. Such a body cannot swim on its own, for there is no load or cell to counter balance the torque it would exert on the surrounding fluid during rotation, so it must be motionless as seen in a fixed, laboratory frame. However, this body may be represented as an active helical body propagating a wave with velocity $c$ plus a rigid body rotation that contributes a wave with speed $-c$. That the body frame may be chosen arbitrarily allows for such an ambiguity in the definition of the swimming speed, but in the laboratory frame this ambiguity disappears. 

A similar argument can be used to show that $c$ has no bearing on the swimming speed of a superhelical flagellum in the laboratory frame. For a superhelical flagellum, the propagation of the major wave can be defined as a rigid body rotation of the entire superhelical structure about the longitudinal axis ($z$-axis) in the body frame. The apparent rotation rate is the sum of the rotation caused by the active propagation of the major wave and the rotation caused by fluid reaction to maintain the force-free and torque-free conditions. For swimming without a head, only the apparent rotation rate is important because the fluid forces and torques are functions of the apparent rotation rate alone (not the absolute value of the rotation rate due to the active major wave propagation). Any alternation of the major wave speed is accompanied by a corresponding change in the rotation rate caused by the fluid reaction, resulting in the same apparent rotation rate and the same force and torque balances. Therefore, the kinematics and dynamics do not depend on the absolute value of the major wave speed $c$ (or $c_m$). In other words, the major wave speed does not contribute to swimming.

For the case of swimming with a head, the fluid reaction will not only rotate the flagellum but also the sperm head, which creates extra torques and perturbs the original force and torque balances. Altering the absolute value of the major wave speed in this case theoretically will affect the swimming velocities even in the laboratory frame, because it changes the relative portion of the rotation rate caused by the fluid reaction and hence the value of the extra torque from the sperm head. However, since the sperm head is slender and short compared with the overall length of the flagellum of actual insect spermatozoa, the extra resistant forces and torques created should be insignificant. Therefore, we suggest that the propagation of the major wave contributes very little to the propulsion (none in the case of head-less swimming), and that it is the minor wavespeed which is primarily responsible for the propulsion. Note also that the hydrodynamic efficiencies in \S\ref{sec:efficiency} are independent of the value of $c$. Therefore, from a hydrodynamic efficiency point of view, there is no advantage or disadvantage to actively propagate a major wave. However, there are other energy costs not taken into account here (for example, work done to produce the sliding of microtubules within the flagellum) in actively propagating a major wave. Therefore, we speculate that it might be energetically more favorable for a doubly-helicated organism to propagate only a minor wave.

 \begin{figure*}
\begin{center}
\includegraphics[width=0.85\textwidth]{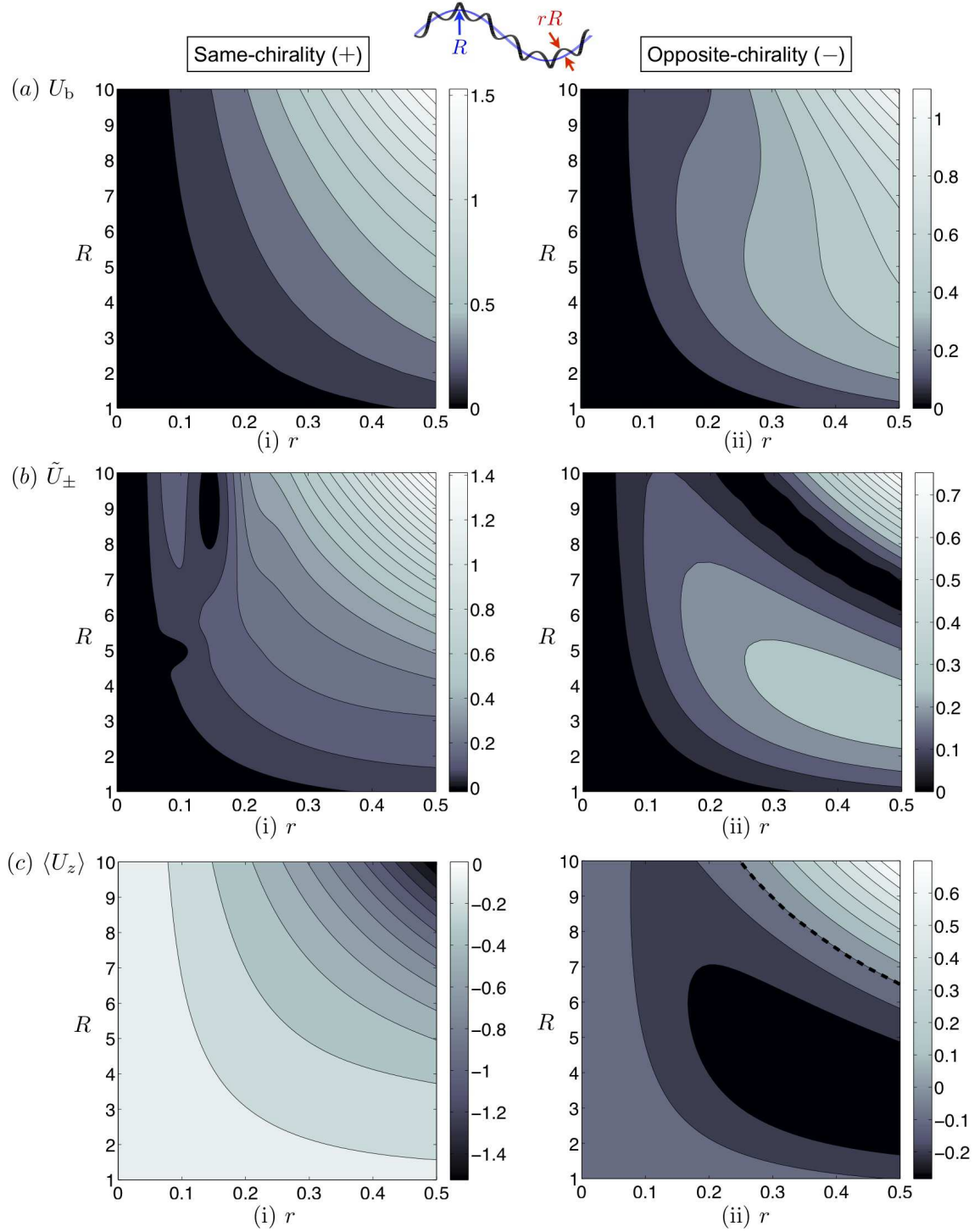}
\end{center}
\caption{\footnotesize  Parametric study of the dependence of the average swimming speed in the body fixed frame $U_\text{b}$ (a(i)\&(ii)) and in the laboratory frame $\tilde{U}_{\pm}$ (b(i)\&(ii)) as a function of the dimensionless parameters $R$ and $r$ (see the schematic for geometrical illustration). Panels c(i) \& c(ii) show the average swimming velocity in the $z$-direction in the body frame, $\langle U_z \rangle$, as a function of $R$ and $r$. The panels on the left (right) refer to the same- (opposite-) chirality configuration. The dotted line in panel c(ii) represent the contour of $\langle U_z \rangle = 0$. Geometric data of \textit{Culicoides melleus} spermatozoa are used for other fixed parameters.}
\label{fig:Rr}
\end{figure*}

\subsubsection{The effect of $R$ and $r$}
We now examine the effect of geometrical dimensionless parameters $R$ and $r$, keeping all other parameters of \textit{Culicoides melleus} spermatozoa fixed. $R$ is the dimensionless major wave amplitude and $r$ is the ratio of minor to major wave amplitude, hence the minor wave amplitude is given by $rR$ (see the schematic in Fig.~\ref{fig:Rr}). In general, one expects the propulsion speed to increase with the wave amplitude for simple geometries. However, for a superhelical structure, the dependence of the average swimming speed on the minor wave amplitude displays interesting behavior. Physically, both the propulsive force and the bulkiness of the structure are varied upon changing the major or minor wave amplitudes. The competition between these factors and the coupling between kinematics in different directions create interesting geometric dependencies of the average swimming speed. We illustrate this by observing the average swimming speed under different frames of reference.

First, we look at the time-averaged swimming velocity in the body frame $U_\text{b} = |[ \langle U_x(t) \rangle,  \langle U_y(t) \rangle, \langle U_z(t) \rangle]|$. As shown in Fig.~\ref{fig:Rr}a(i), for the same-chirality configuration the average swimming velocity grows monotonically with $R$ and $r$. Note that, keeping other parameters fixed, increasing the value of $R = A_M k_m$ for a fixed $r = A_m / A_M$ geometrically means that both the major $A_M$ and minor $A_m$ amplitudes are increased simultaneously by the same proportion. There is a competition between the increase in the overall hydrodynamic resistance due to the increased bulkiness (correlated with increases in $A_M$ and $A_m$) and an enhanced propulsive force (correlated with an increase in $A_m$). In the case of the same-chirality configuration, the latter effect dominates. However, for the opposite-chirality configuration, as shown in Fig.~\ref{fig:Rr}a(ii), non-monotonic variations in the swimming speed with the major amplitude $R$ are observed for certain values of $r$, the ratio of the minor to major wave amplitudes.

 \begin{figure*}
\begin{center}
\includegraphics[width=1\textwidth]{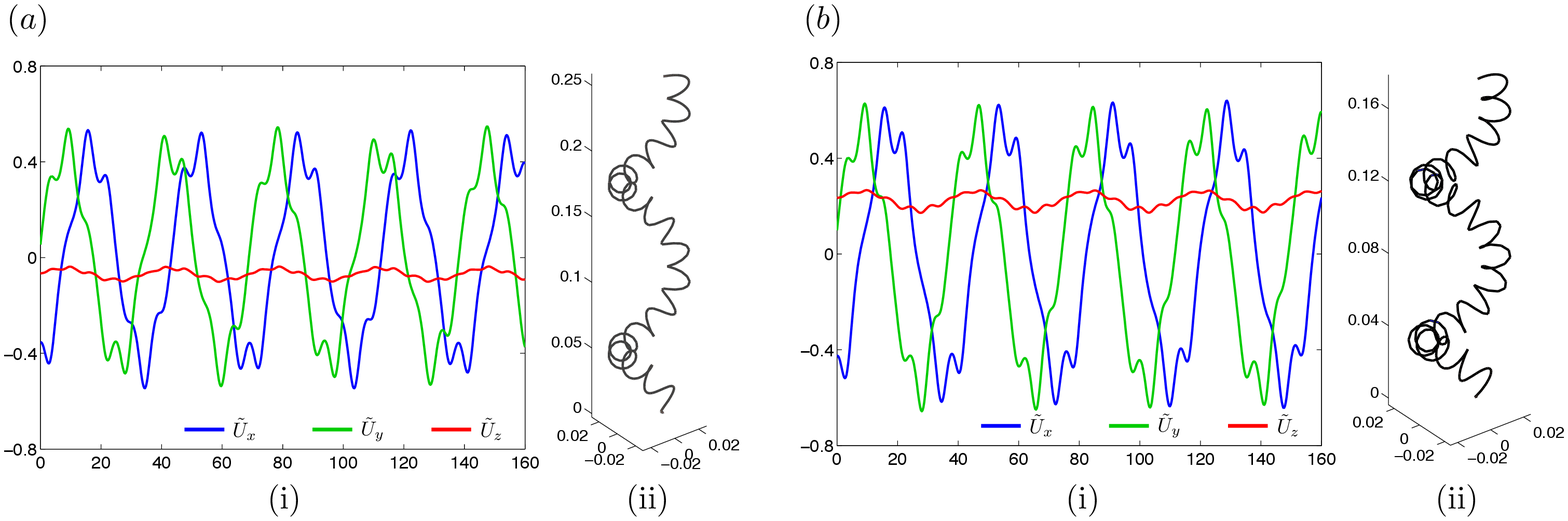}
\end{center}
\caption{\footnotesize  Three-dimensional swimming velocities in the laboratory frame ((a)i \& (b)i) of two opposite-chirality superhelical swimmers, and their corresponding geometries ((a)ii \& (b)ii). (a) $R=7$, $r=0.4$, and $U_z$ is negative, while the wave propagation is towards the positive $z$-direction; (b) $R=9$, $r=0.4$, and $U_z$ is positive, while the wave propagation is towards the positive $z$-direction. Geometric data of \textit{Culicoides melleus} spermatozoa are used for other parameters.}
\label{fig:bidirection}
\end{figure*}

Since the swimming kinematics are three-dimensional, variations of the geometric parameters affect swimming velocities and rotational rates in all directions. In particular, a large propulsion speed in the body frame does not necessarily imply a large net propulsion speed in the laboratory frame. The mean swimming speeds in the laboratory frame for both chirality configurations are shown in Fig.~\ref{fig:Rr}b(i-ii). The coupling between the swimming kinematics in all directions produces more complicated variations in the propulsion speed as a function of $R$ and $r$. Non-monotonic behaviors are observed in both cases.

We have already noted that some superhelical swimmers propel themselves surprisingly in the same direction as the wave propagation, unlike for planar or single helical wave propulsion. For the range of parameters explored in this paper, this direction reversal is found to occur only in the opposite-chirality configuration for sufficiently large $R$ and $r$ (Fig.~\ref{fig:Rr}c(ii)), while the average propagation velocity $\langle U_z \rangle$ is always negative for the same-chirality configuration (i.e.~the swimming direction is always opposite to the direction of the propagating wave) (Fig.~\ref{fig:Rr}c(i)). In Fig.~\ref{fig:bidirection}, we show the detailed swimming velocities of two opposite-chirality superhelical swimmers, where one of them has its longitudinal propulsion in the opposite direction relative to the wave propagation ($R=7$, $r=0.4$, Fig.~\ref{fig:bidirection}a), and the other in the same direction as the wave propagation ($R=9$, $r=0.4$, Fig.~\ref{fig:bidirection}b). It is intriguing that a small change in the geometry (see the corresponding superhelices in Fig.~\ref{fig:bidirection}a(ii) \& b(ii)) can lead to a swimming direction reversal. We argue in the next section (\S\ref{sec:SBTvsRFT}) that this transition is related to the hydrodynamic interaction between distinct parts of the superhelical flagellum.

\subsection{\label{sec:SBTvsRFT}Comparison between slender body theory and resistive force theory}

 \begin{figure*}
\begin{center}
\includegraphics[width=0.83\textwidth]{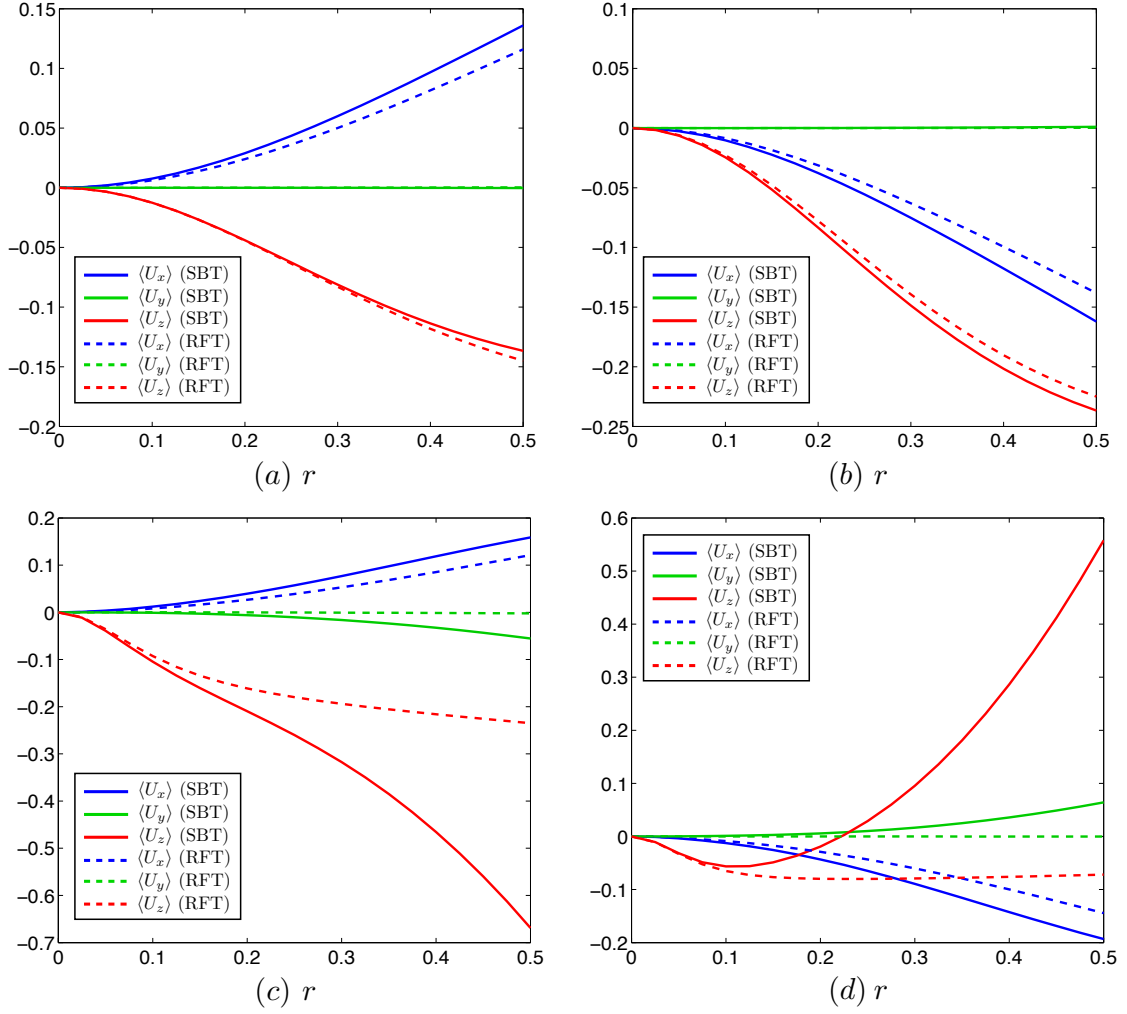}
\end{center}
\caption{\footnotesize  Comparison between the resistive force theory (RFT) (dashed lines) and the non-local slender body theory (SBT) (solid lines), using geometric data of \textit{Culicoides melleus} spermatozoa (a) \& (b), and \textit{Tenebrio molitor} spermatozoa (c) \& (d). The panels on the left (right) refer to the same- (opposite-) chirality configuration.}
\label{fig:SBTRFT}
\end{figure*}

In order to consider the relative importance of nonlocal hydrodynamic interactions in the swimming of superhelices, we now compare our results to those obtained using the more commonly used local drag model (eq.~\eqref{eqn:SBT}, but neglecting the non-local term $\mathbf{K}$). The local drag model (so-called resistive force theory) ignores hydrodynamic interactions between distinct parts of the curved flagellum, and is expected to work well \cite{wiggins98, yu06, kruse07} for simple geometries where different parts of the body are sufficiently well separated. However, for more complicated geometries, the local drag model may not capture even the correct qualitative features. Recently, Jung \textit{et al.} \cite{jung07} studied the rotational dynamics of opposite-chirality superhelices and found a qualitative discrepancy (the rotational direction of the superhelix when being towed in a viscous flow) between the experimental results and the predictions from the resistive force theory. Other recent works have also shown the inadequacy of the local drag model \cite{cw09,sl11}.

Figure~\ref{fig:SBTRFT} shows the swimming speeds computed with both non-local slender body theory and resistive force theory, using two sets of geometrical data: \textit{Culicoides melleus} (Fig.~\ref{fig:SBTRFT}a (same-chirality) \& b (opposite-chirality)), and \textit{Tenebrio molitor} (Fig.~\ref{fig:SBTRFT}c (same-chirality) \& d (opposite-chirality)). In these two cases, all other parameters are fixed but the amplitude ratio $r$ is varied from 0 to 0.5. Increasing $r$ complicates the swimmer geometry, and the hydrodynamic interactions are expected to be more significant for large $r$. For the case of \textit{Culicoides melleus} (Fig.~\ref{fig:SBTRFT} a \& b), we see good (even quantitative) agreement between the results form the slender body theory (solid lines) and the resistive force theory (dashed lines). However, for the case of \textit{Tenebrio molitor} (Fig.~\ref{fig:SBTRFT} c \& d), which has larger values of $R$ and $K$, the deviation between the two models becomes significant for the same-chirality configuration (Fig.~\ref{fig:SBTRFT}c) as $r$ increases. There are even qualitative discrepancies: in the opposite-chirality configuration (Fig.~\ref{fig:SBTRFT}c) for large values of $r$, the resistive force theory fails to capture the transition in the sign of $\langle U_z \rangle$ predicted by the slender body theory. Since no such transition is found when the hydrodynamic interactions are ignored, this transition may be attributed to nonlocal hydrodynamic interactions of the body with itself. In general, and perhaps unsurprisingly, we have shown that the local drag model breaks down when the geometry of the structure is sufficiently intricate.

\subsection{Asymptotic analysis}
In the locomotion of some species of spermatozoa, the minor wave amplitude is much smaller than the major wave amplitude (see Tables~\ref{table:geometry} \&~\ref{table:dimensionless}), which motivates us to perform an asymptotic analysis for $r \ll 1$. Such an asymptotic analysis linearizes the problem geometrically and allows the nonlinear effects to be taken into account order by order, making the problem more amenable to mathematical analysis. However, even in the asymptotic consideration, three-dimensional force and torque balances do not yield tractable analytical results. Hence, in the spirit of Chwang and Wu \cite{chwang71}, we perform the force and torque balances only in the longitudinal ($z$-) direction using the resistive force theory, which is expected to be at least qualitatively correct in the asymptotic limit $r \ll 1$. The local force $f_z$ and torque $m_z$ may be expressed as
\begin{align}
f_z &= c_{U_zF_z}(t,s) U_z+c_{\Omega_zF_z}(t,s) \Omega_z+c_{F_z}(t,s)\label{eq:asymForce},\\
m_z  &= c_{U_zM_z}(t,s) U_z+c_{\Omega_zM_z}(t,s) \Omega_z+c_{M_z}(t,s)\label{eq:asymTorque},
\end{align}
where the coefficients are determined analytically from Eq.~\ref{eqn:SBT} (without the non-local operator) and the geometry of the swimmer (Eqs.~\ref{eqn:deformationX} to~\ref{eqn:deformationZ}). We consider regular perturbation expansions in $r$ for every term in Eqs.~\ref{eq:asymForce} and~\ref{eq:asymTorque}, and enforce the force-free and torque-free conditions order by order. A non-zero time-averaged swimming velocity enters at  $O(r^2)$. The leading order mean swimming velocities for the cases of same- and opposite-chirality configurations read
\begin{align}
\tilde{U}_{\pm} \sim & \  \frac{R^2 r^2}{2 L^2 \left(1+R^2 K^2\right)^{3/2}} \times \Bigl\{ \left(2-2L^2\right) (\xi -1) \notag \\
& \ +R^2 K^2\left[2 (\xi -1\pm K)-L^2 (2 \xi -2\pm K)\right] \notag \\
& \ -2 \left[\xi -1+R^2 K^2(\xi -1\pm K)\right] \cos L \Bigr\}  \label{eq:asym}, 
\end{align}  
\normalsize
where $\xi=\xi_{\perp}/\xi_{\parallel}$ is the ratio of the drag coefficients in the normal direction to the longitudinal direction ($\xi_{\parallel} = 4\pi/c_0, \xi_{\perp}=8\pi/(2+c_0)$). For very slender filaments, $\xi \approx 2$. We note that the substitution $K \rightarrow -K$ in Eq.~\ref{eq:asym} converts the same-chirality speed $\tilde{U}_{+}$ to the opposite-chirality speed $\tilde{U}_{-}$. The propulsion speed has a quadratic dependence on the minor wave amplitude for small $r$, which is also found in planar \cite{taylor1} and helical \cite{taylor2} geometries. 

\begin{figure*}
\begin{center}
\includegraphics[width=0.85\textwidth]{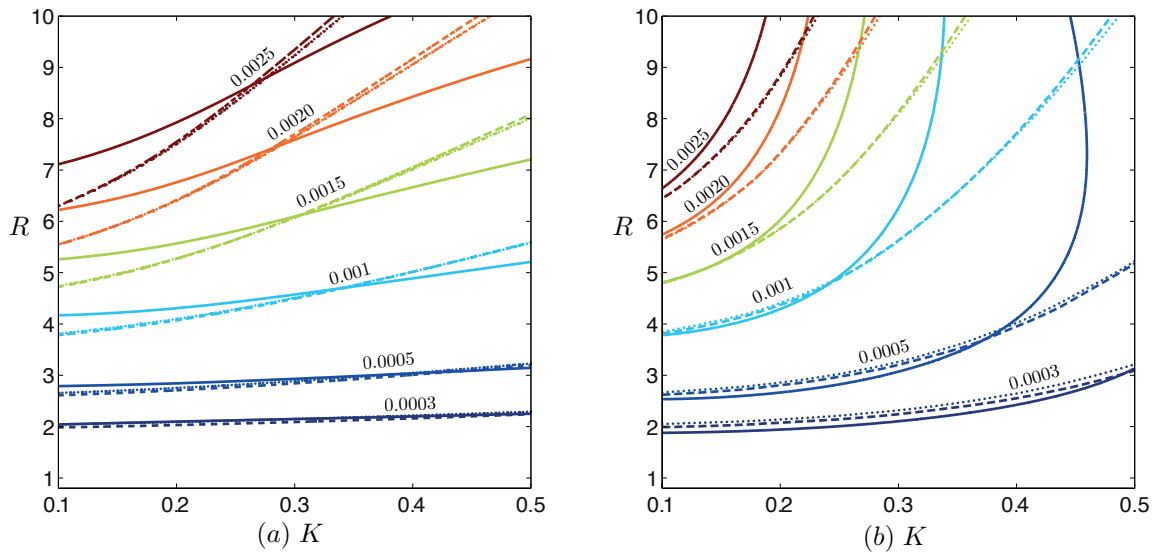}
\end{center}
\caption{\footnotesize  Comparison of the asymptotic results (dotted lines) with the finite-amplitude simulations using the resistive force theory (dashed lines) and the non-local slender body theory (solid lines). Shown are the contour lines of the average swimming velocity $\tilde{U}_{\pm}$ for the (a) same-chirality, and (b) opposite-chirality configuration for a small minor to major wave amplitudes ratio of $r=0.01$.}
\label{fig:asym}
\end{figure*}

These asymptotic results (Eq.~\ref{eq:asym}) are compared to the finite amplitude simulations (both the resistive force theory and the slender body theory predictions) in Fig.~\ref{fig:asym}, for a very small amplitude ratio of $r=0.01$ and a slenderness ratio of $\epsilon=1/1000$. There is excellent agreement between the asymptotic results and the finite amplitude resistive force theory simulations. The discrepancy between the non-local slender body theory and the local drag model highlights the importance of non-local hydrodynamic interactions for such organisms.

We conclude by pointing out an intriguing theoretical curiosity, that drag anisotropy ($\xi \neq 1$) is not required for superhelical swimming: setting $\xi=1$ in Eq.~\ref{eq:asym}, non-zero mean propulsion velocities (of equal magnitude but opposite signs) are obtained for both chirality configurations,
\begin{align}
\tilde{U}_{\pm} (\xi=1) &= \mp \frac{R^4 K^3 (L^2-2+2 \cos L)}{2 L (1+R^2 K^2)^{3/2}} r^2.
\end{align}
Note that Becker \textit{et al.}'s \cite{becker} argument of the requirement of drag anisotropy for locomotion is only true for inextensible swimmers and does not apply here. The superhelical kinematics described (Eqs.~\ref{eqn:deformationX} to~\ref{eqn:deformationZ}) are only possible when extensibility is allowed; the minor helix is built upon another curved structure (the major helix) and local extension and contraction is implied in the wave kinematics. When extensibility is permitted, the relaxation of the drag anisotropy requirement has been recently shown \cite{pak}. That the swimming speed is non-zero is due to intrinsic variations in length (and hence drag) embedded in the curved geometry of the superhelices. A minor helix built upon a major helix has relatively shorter lengths in the regions closer to the longitudinal axis, creating an overall imbalance of hydrodynamic drag even in the isotropic drag case ($\xi=1$). A similar example is a toroidal helix (a helix built upon a circle), which is an idealized model studied recently for dinoflagellates \cite{nguyen}. We expect that the propagation of a wave along a toroidal helix should also require extensibility, and that propulsion is still possible even without drag anisotropy \cite{pak}.

\section{\label{sec:discussion}Discussion}

In this paper we have studied a morphologically interesting double-wave structure exhibited by various insect spermatozoa. The construction of such spermatozoa is considerably more complex than those for flagella which exhibit simpler planar or helical waves: the flagellum does not only have a more complicated 9+9+2 arrangement of microtubules but also mitochondrial derivatives and accessory bodies running along the axoneme \cite{werner08}. We have mathematically idealized the double-wave structure as a superhelical structure and presented a hydrodynamic study on superhelical swimming. The available data is primitive and sparse; nevertheless, we consider the agreement between experimental measurements and the theory explored herein to be quite reasonable. Through numerical experiments, we have found that the major wave speed has little contribution to propulsion when the sperm head is small, as is the case for insect spermatozoa. When there is no sperm head, the propulsion speed is independent of the major wave speed and depends entirely upon the minor wave speed. We have also explored the dependence of the propulsion speed on the dimensionless major wave amplitude $R$ and the ratio of the minor to major wave amplitudes $r$ (Fig.\ref{fig:Rr}), and counter-intuitive behaviors have been found for the opposite-chirality configuration. In particular, we have found that propulsion and wave propagation can occur in the same direction for superhelices in the opposite-chirality configuration.

The present study suggests that the major wave has negligible influence on the motility of a superhelical swimmer. This finding favors the recent hypothesis by Werner \textit{et al.} \cite{werner02} that the major helical wave is a static (non-propagating) structure; the minor wave structure is solely responsible for the motility, and the apparent major wave propagation is simply due to the passive rotation of the entire geometry (see \S\ref{sec:introduction}). However, in the study by Baccetti \textit{et al.} \cite{baccetti89} on the motility of \textit{Ceratitis capitata}, they have adopted the same experimental techniques as in Gibbons \textit{et al.} \cite{gibbons85}, which distinguished the rolling frequency from the apparent beat frequency. In their work, the rolling frequency was measured by stroboscopic observation of the eccentrically attached sperm head, and the flagellar beat frequency was measured by the same means with the sperm head adhered to the bottom of the observation dish. Using this method, the major wave speed measured should be taken as an active propagation speed. We are not in a position to provide a definite answer on whether or not the major wave propagates actively in actual insect spermatozoa. However, according to the present study, we suspect that there might be some biological reasons, other than motility, for the major wave to propagate actively. We also do not know if the propagation of the major wave is a biological prerequisite for the propagation of the minor wave. Further biological studies are required to answer these questions.

It is illuminating to compare the superhelical swimming studied here to regular helical swimming. The swimming trajectories are qualitatively different in the two cases: in regular helical swimming, the trajectory is a regular helix, whereas in superhelical swimming the trajectory is doubly-helicated. In addition, a regular helical flagellum cannot swim on its own; a sperm head is required to swim (since the absence of a sperm head renders the deformation of the regular helix a rigid body motion). In contrast, the propagation of superhelical waves along a flagellum is in general not equivalent to rigid body motion, and hence ``head-less" swimming is possible. This might help explain the presence of only a small slender sperm head in insect spermatozoa. In other words, superhelical swimming can be viewed as an alternative mechanism to regular helical swimming when only a small sperm head is available.

Finally, a non-local slender body theory was used in this work and compared with a simpler and widely used local drag model. We showed that the resistive force theory failed to capture even the qualitative features of the swimmer when the geometry becomes complicated. The results suggest that further hydrodynamic studies on superhelical structures require more advanced models than the local resistive force theory, such as the slender body theory employed here. Optimization with respect to the efficiency of the swimmer taking into account of the viscous dissipation, and other energy costs due to bending and internal sliding of filaments \cite{sl10}, which have been neglected in this work, will be interesting for future work.

{\begin{acknowledgements}
This work was funded in part by the US National Science Foundation (Grant CBET-0746285 to E.~L.) and the Croucher Foundation (through a scholarship to O.~S.~P.). Useful conversations with  Dr.~Michael Werner and Dr.~Anne Swan are gratefully acknowledged.
\end{acknowledgements}}

\end{document}